\documentclass[12pt,aps,preprint,nofootinbib,superscriptaddress,nobalancelastpage]{revtex4}
\usepackage{amsmath,amssymb}
\usepackage{graphicx}
\usepackage{multirow}
\usepackage{hyperref}
\usepackage{epsfig}
\usepackage{amsfonts}
\usepackage{bbm}
\usepackage{mathrsfs}
\usepackage{color}

\newcommand{\be}{\begin{equation}}
\newcommand{\ee}{\end{equation}}
\newcommand{\bea}{\begin{eqnarray}}
\newcommand{\eea}{\end{eqnarray}}

\newcommand{\ba}{\begin{array}}
\newcommand{\ea}{\end{array}}
\newcommand{\balg}{\begin{align}}
\newcommand{\ealg}{\end{align}}

\newcommand{\fig}{Fig.}

\newcommand{\tab}{Tab.}
\newcommand{\BR}{{\rm BR}}
\newcommand{\eg}{\emph{e.g.}}
\newcommand{\cf}{c.f.}
\newcommand{\ie}{\emph{i.e.}}
\renewcommand{\sec}{Sec.}
\newcommand{\Ref}{Ref.}
\newcommand{\kton}{kt}

\newcommand{\mycaption}[1]{\caption{#1}}

\newcommand{\lsim}
{\raise0.3ex\hbox{$\;<$\kern-0.75em\raise-1.1ex\hbox{$\sim\;$}}}
\newcommand{\gsim}
{\raise0.3ex\hbox{$\;>$\kern-0.75em\raise-1.1ex\hbox{$\sim\;$}}}

\begin{document} 

\title{Neutrino Probes of the Nature of Light Dark Matter}

\author{Sanjib Kumar Agarwalla}
\email[]{Sanjib.Agarwalla@ific.uv.es}
\affiliation{Instituto de F\'{\i}sica Corpuscular, CSIC-Universitat de Val\`encia, \\
Apartado de Correos 22085, E-46071 Valencia, Spain}

\author{Mattias Blennow}
\email[]{blennow@mppmu.mpg.de}
\affiliation{Max-Planck-Institut f\"ur Physik
(Werner-Heisenberg-Institut), F\"ohringer Ring 6, 80805 M\"unchen,
Germany}

\author{Enrique Fernandez Martinez}
\email[]{enfmarti@cern.ch}
\affiliation{CERN Physics Department, Theory Division \\ CH-1211 Geneva 23, Switzerland}

\author{Olga Mena}
\email[]{omena@ific.uv.es}
\affiliation{Instituto de F\'{\i}sica Corpuscular, CSIC-Universitat de Val\`encia, \\
Apartado de Correos 22085, E-46071 Valencia, Spain}

\begin{abstract}
{Dark matter particles gravitationally trapped inside the Sun 
may annihilate into Standard Model particles, producing a flux 
of neutrinos. The prospects of detecting these neutrinos in future
multi-\kton{} neutrino detectors designed for other physics searches 
are explored here. We study the capabilities of a 34/100~\kton{} liquid 
argon detector and a 100~\kton{} magnetized iron calorimeter detector. 
These detectors are expected to determine the energy and 
the direction of the incoming neutrino with unprecedented precision allowing for tests of the dark matter nature at very low dark matter masses, in the range of $10-25$~GeV. By suppressing the atmospheric background with angular cuts, these techniques would be sensitive to dark matter - nucleon spin-dependent cross sections 
at the fb level, reaching down to a few ab for the most favorable annihilation channels and detector technology.} 
\end{abstract}



\preprint{CERN-PH-TH/2011-118, EURONU-WP6-11-35, IFIC/11-24, MPP-2011-56}

\maketitle

\section{Introduction}
\label{sec:introduction}

 A plethora of cosmological and astrophysical measurements over the last decades has tested the validity of the Standard Model of Big Bang cosmology (a spatially flat Friedman--Robertson--Walker model) at an unprecedented level of precision. Current observations point to a flat universe in which the mass-energy content includes only 5~\% of ordinary matter (baryons) while  22~\% is a mysterious non-baryonic dark matter~(DM) component, whose properties are largely unknown. If it can be identified with some relic, this should be a massive, neutral and sufficiently long-lived particle. A popular hypothesis, realized in many extensions of the Standard Model of particle physics~(SM), including supersymmetry~\cite{Jungman:1995df} or extra dimensional models~\cite{Servant:2002aq,Cheng:2002ej}, is that the DM is dominated by a single species of weakly interacting massive particle~(WIMP). Such WIMPs could have masses ranging from several GeV to dozens of TeV, see \Ref~\cite{Bertone:2004pz} for a review. Direct DM searches look for the recoil energy of target nuclei due to interactions with the DM particles. WIMP scattering on nuclei may happen via spin-dependent interactions (proportional to $J(J+1)$, rather than to the number of nucleons) or via spin-independent scattering (which grow with $A^2$ and therefore increase dramatically with the mass of the target nuclei). The spin-independent cross section dominates over spin-dependent scattering in current direct DM detection experiments, which use heavy atoms as targets, setting therefore very strong constraints on spin-independent interactions. For DM capture in the Sun, however, the distinction is not so strong as in direct detection searches since a large fraction of the Sun's mass is in the form of hydrogen, for which there is no enhancement of the spin-independent cross section. In the following, we will present our results in terms of spin-dependent DM scattering cross section. However, in practice, similar neutrino fluxes would be obtained with spin-independent cross section around two orders of magnitude smaller due to its enhancement for the heavier nuclei present in the Sun (see, \eg, \Ref~\cite{Kappl:2011kz}). Thus, all the results presented here for spin-dependent scattering cross sections can be roughly translated for the spin-independent case with an improvement of around two orders of magnitude.  

Although there is no confirmed observation of the passage of WIMPs through a detector, a long standing positive signal in the form of an annual modulation has been  reported by the DM search experiment DAMA/LIBRA~\cite{Bernabei:2010mq}. This annual modulation is expected in the scattering of DM particles off nuclei in the detector due to the Earth's annual motion around the Sun, that modifies the relative velocity of the Earth and the DM halo. The DAMA/LIBRA collaboration has reported the observation of 13 such cycles, compatible with the expected phase due to the Earth's orbit, at $8.9\sigma$. This observation is, however, not confirmed by other direct search experiments, such as Xenon~\cite{Aprile:2011hi,Angle:2011th} and CDMS~\cite{Ahmed:2010wy}. Tension between the two datasets remains even when interpreting the DAMA/LIBRA results as scatterings of DM particles off the Na nuclei\footnote{The DAMA detector exploits very pure sodium iodide crystals and DM scatterings off the I nuclei would instead point to heavier DM masses.}, which would point to light DM masses $\sim 10$~GeV with a spin-independent scattering cross section $\sigma \sim 0.2$~fb close to the threshold of the Xenon and CDMS sensitivities. On the other hand, the CoGeNT experiment has also reported an observed excess of events over the expected background~\cite{Aalseth:2010vx}. As in the DAMA/LIBRA case, the shape of the excess would also be consistent with scatterings of light DM $\sim 10$~GeV, although with somewhat smaller spin-independent scattering cross section $\sigma \lesssim 0.1$~fb. The similarity of the DAMA/LIBRA and CoGeNT results is very intriguing, although there is some tension between the preferred value of the cross section and there is controversy on whether the two DM hints can be reconciled in the simple spin-independent elastic scattering picture, with analyses pointing to a combined allowed region around $\sim 7$~GeV and $\sigma \sim 0.2$~fb~\cite{Hooper:2010uy} or disfavouring such an explanation~\cite{Schwetz:2010gv}. Even more tantalizing is the recent announcement of the CoGeNT collaboration of an annual modulation, similar to the DAMA/LIBRA one, in their observed excess at $2.8\sigma$~\cite{cogent}. Furthermore, the CRESST-II experiment also observes an excess of 32 events to be compared with an expect background of $8.4 \pm 1.4$ in the O band~\cite{Jochum:2011zz}, corresponding to the lowest energies. When interpreted as a DM hint, these events could be accommodated with a mass and cross section in the same region as that favoured by DAMA/LIBRA and CoGeNT~\cite{Schwetz:2010gv}. While the accumulation of hints favouring light DM with masses of $\sim 10$~GeV and a spin-independent cross section $\sigma \sim 0.1$~fb is tantalizing, this region is disfavoured by the null results of the Xenon and CDMS searches. However, as this region is close to the energy threshold, these results are sometimes questioned since small changes in the energy calibration or efficiencies could indeed reopen the window required for the DM interpretation of DAMA/LIBRA, CoGeNT and CRESST. On the other hand, the Xenon and CDMS experiments have a cleaner signal region, given the stronger background suppression achieved. Therefore, the DAMA and CoGeNT excesses could be due to a misinterpreted background component. A very similar situation occurs when interpreting the DAMA/LIBRA signal through a spin-dependent cross section in the pb range and the null results of the PICASSO~\cite{Archambault:2009sm}, COUPP~\cite{Behnke:2010xt} and SIMPLE~\cite{Felizardo:2011uw} searches. 

Clearly, more experiments sensitive to these low DM mass regions are needed in order to settle the present tensions and shed light on the DM puzzle. In this context, DM indirect detection methods offer a very attractive complementary probe. They aim to detect the gamma-ray, positron, anti-proton and neutrino fluxes produced in DM annihilations in regions with dense DM accumulations, such as the Galactic center. In the case of neutrinos, the higher densities of DM expected to be gravitationally trapped in celestial bodies, such as the Sun~\cite{Silk:1985ax,Krauss:1985ks} or the Earth~\cite{Freese:1985qw,Krauss:1985aaa} offer a very appealing alternative since, contrary to positrons or gamma-rays, neutrinos can escape the solar interior providing a snapshot of the interaction taking place. These indirect searches, provide a very complementary tool to direct detection probes since, unlike them, they will not only provide a measurement of the DM mass and scattering cross section, but also information on the different annihilation branching ratios, depending on the particular channel searched for. Bounds on the WIMP mass and cross section from the water Cerenkov (WC) neutrino detector Super-Kamiokande (Super-K) have been obtained for different annihilation channels, see \Ref~\cite{Desai:2004pq,Niro:2009mw,Kappl:2011kz}. A lot of effort has also been devoted to using neutrino telescopes to detect neutrinos from DM annihilation processes~\cite{Boliev:1995xz,Barger:2001ur,Ackermann:2005fr,Halzen:2005ar,Abbasi:2009uz,Loucatos:2010zz}. However, these neutrino telescopes are mostly sensitive to $\nu_\mu$ and $\bar\nu_\mu$ events, and their energy threshold is $\sim 5-10$ GeV in the muon energy (which is translated into $\sim 15-20$~GeV neutrino energy). Other neutrino detectors might be necessary to test the DM mass region indicated by DAMA/LIBRA and CoGENT via indirect searches. Even if Super-K is sensitive to low DM masses, the energy resolution achieved by the former detector is not as good as for some future facilities. Future neutrino detectors, such as WC, liquid argon (LAr), liquid scintillator, or magnetized iron neutrino detector (MIND) facilities, might be constructed for other physics measurements (such as neutrino oscillation studies or core collapse supernova neutrino physics) and could be exploited for indirect DM detection.

As first proposed in \Ref~\cite{Mena:2007ty}, future MIND-like detectors currently under development, such as India-based Neutrino Observatory~(INO), could provide an excellent energy and angular resolution to neutrinos from WIMP annihilations in the Sun. In such detectors, the expected degeneracies among the DM mass, annihilation branching ratios, and elastic cross section with nuclei could be resolved, shedding much light on the DM nature. Other detection techniques originally proposed and constructed for long baseline and supernova neutrino studies have also been examined in this context, such as future LAr detectors~\cite{Bueno:2004dv} as well as current and future liquid scintillator detectors~\cite{Kumar:2011hi}.

Here we extend the work of \Ref~\cite{Mena:2007ty} by considering both lower DM masses ($\sim 10$~GeV) and additional annihilation channels to the $b\bar{b}$ and $\tau^-\tau^+$ previously studied. We also consider not only future MIND detectors, but also proposed future LAr facilities, such as GLACIER. Both techniques are expected to achieve unprecedented energy and angular reconstruction capabilities at $\sim$~GeV neutrino energies.
 
 The structure of this paper is as follows: Section~\ref{sec:detectors} describes the MIND and LAr detector technologies exploited here. The neutrino fluxes and event rates from WIMP annihilations in the Sun, as well as backgrounds to our signal from atmospheric neutrino events in these two detector facilities are presented in \sec~\ref{sec:fluxesandrates}. In \sec~\ref{sec:reconstruction}, we present the DM nature reconstruction exercise performed here by means of  a Markov Chain Monte Carlo (MCMC) analysis for several {light} DM particle masses. We summarize and draw our conclusions in \sec~\ref{sec:summary}.

\section{Large Neutrino Detectors}
\label{sec:detectors}

\begin{figure}
\centering
\includegraphics[width=0.5\textwidth]{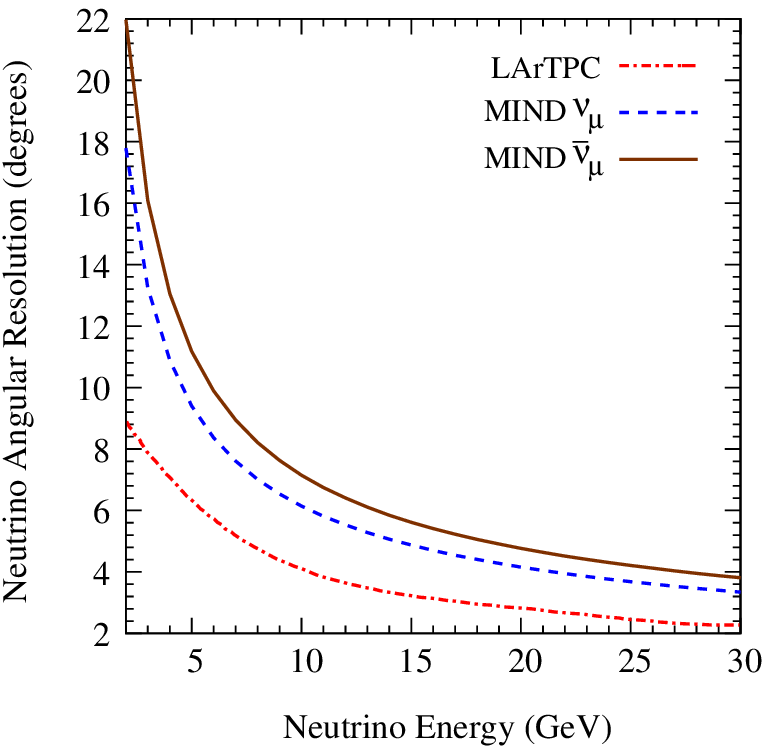}
\mycaption{The angular resolutions for LArTPC and MIND considered in this study as a function of the true neutrino energy. These curves have been derived from~\cite{Bueno:2004dv,ThesisLaing,Bari:2003bt}.
\label{fig:resolution}}
\end{figure}

Large underground neutrino detectors~\cite{Hirata:1987hu,Fukuda:1998mi,Ahmad:2002jz,:2008ee,Arpesella:2008mt,Ahn:2006zza,Adamson:2008zt} have been very successful in observing neutrinos from several natural and artificial 
sources and have provided one of the few evidences we have of physics beyond the SM, as well fundamental results in astrophysics~\cite{GonzalezGarcia:2007ib,Strumia:2006db}. Active research is underway worldwide to plan and build the next-generation of large underground neutrino detectors in the 
mass range of 10$^2$ to 10$^3$ kt~\cite{Itow:2001ee,Autiero:2007zj,Raby:2008pd,Lesko:2008zza,Raychaudhuri:2008zz,Datar:2009zz,Rubbia:2010zz}, which will play a crucial role 
in the upcoming era of precision neutrino science. 
The physics potential of such large detectors is extremely broad. One of the major aims is the 
search for proton decay~\cite{Pati:1973rp}, which provides a direct test of Grand Unified Theories. These large neutrino detectors are also essential for the future accelerator based long baseline neutrino oscillation experiments to explore mixing and CP violation in the lepton sector~\cite{Rubbia:2010fm,Barger:2007yw,Bandyopadhyay:2007kx}. Applications in astrophysics include precise measurements of supernova neutrinos and the search for new sources of astrophysical neutrinos~\cite{Autiero:2007zj,Wurm:2011zn}. In this context, the possibility of observing the neutrinos originated from the annihilation of DM particles in the center of the Sun~\cite{Desai:2004pq,Bueno:2004dv,Mena:2007ty,Raby:2008pd} and to reconstruct the annihilation process so as to probe the nature of DM is particularly tantalizing and will be the main goal of our study. These detectors are expected to measure the energy and the direction of the initial neutrino with very high precision which allows to search for DM particles in the mass range of $5-50$~GeV, which cannot be probed effectively with large neutrino telescopes such as IceCube~\cite{Abbasi:2009uz}. 

Currently, four different technologies for large underground neutrino detection are being investigated 
worldwide: WC, unsegmented liquid scintillator, LAr time projection chamber (LArTPC), and MIND. Below we discuss in detail the LArTPC and MIND alternatives, which can provide the best performance for the searches under study. Indeed, the energy and angular resolution capabilities of WC and liquid scintillator detectors degrade rapidly beyond the quasielastic regime and thus makes them unsuited for the multi-GeV energies under consideration. The assumed characteristics of the detectors used in our numeric simulations are summarized in \tab~\ref{tab:detector}.
\begin{table}[t]
\begin{center}
\begin{tabular}{|l|c|c|} \hline
\multicolumn{1}{|l|}{{\multirow{2}{*}{Detector characteristics}}}
& \multicolumn{1}{|c|}{{MIND}}
& \multicolumn{1}{|c|}{{LArTPC}}
\\
& (Only $\mu^{\pm}$) & (Both $\mu^{\pm}$ \& $e^{\pm}$) \\
\hline
Fiducial mass & 100~\kton{} & 34/100~\kton{} \\
\hline
$\nu$ energy threshold & 2 GeV & 2 GeV \\
\hline
{Detection efficiency ($\epsilon$)} & 
Fig. 6.21 (Top Panels)  & 100~\% for $\mu^{\pm}$ \\
& of {\Ref~\cite{ThesisLaing}} & 80~\% for $e^{\pm}$ \\
\hline
$\nu$ energy resolution ($\delta E$) & $0.15 E$ & $0.15 \sqrt{E/\rm 1~GeV}$~GeV \\
\hline
$\nu$ angular resolution ($\delta \theta$) & See Fig.~\ref{fig:resolution} & See Fig.~\ref{fig:resolution} \\
\hline
Bin size & 1 GeV & 1 GeV \\
\hline
\end{tabular}
\caption{\label{tab:detector}
Detector characteristics used in our simulations.}
\end{center}
\end{table}

\subsection{Liquid Argon Time Projection Chamber}

A LArTPC~\cite{crubbia} is a modern version of the bubble chamber technology. The working principle of LArTPC
is that by applying an electric field in highly pure argon, free electrons created by the 
passage of an ionizing particle can be drifted over large distances, $\mathcal{O}(\mathrm{m})$, without any distortion.
This enables the possibility to obtain the three dimensional image of the particle track by just reading out the surface 
of the volume. The projections of the track can be used to reconstruct the three dimensional path of the particle with an accuracy 
of a few mm. The viability of the LArTPC technology has been tested by the extensive R\&D 
program, developed by the ICARUS collaboration. Currently, the largest LArTPC ever built, is the ICARUS 
T600 module~\cite{Amerio:2004ze,Varanini:2009zz,Menegolli:2010zz} with a mass of $600\,\mathrm{t}$. 
Based on this technique, a very massive next-generation 100~\kton{} detector 
(GLACIER)~\cite{Rubbia:2004tz,Autiero:2007zj,Rubbia:2010zz} has been proposed under LAGUNA~\cite{laguna}. 
The LBNE physics program in United States is also considering a 34~\kton{} LArTPC as a possible detector candidate for 
DUSEL~\cite{Barger:2007yw,lbne}. In this work, we study the prospects of a 34/100~\kton{} LArTPC detector to search for 
DM WIMPs with 10 full years of data taking. 

Since this technology is still in its R\&D phase, many uncertainties regarding its performance at very large 
mass scale exist. Thus, the detector characteristics that we consider here, are essentially based on the performance 
of the ICARUS T600 detector~\cite{Amerio:2004ze,Varanini:2009zz,Menegolli:2010zz} and on the ongoing estimates of
the LAGUNA~\cite{laguna} and LBNE~\cite{Barger:2006kp,lbne} collaborations.   

In a LArTPC, the muon-electron misidentification probability is almost zero and it offers very high detection 
efficiency for muons and electrons in the GeV energy range, suppressing the neutral current $\pi^{0}$ backgrounds.   
In our study, we consider $\nu_e$, $\bar\nu_e$, $\nu_{\mu}$, and $\bar\nu_{\mu}$ 
charged current (CC) events with a neutrino energy threshold of 2~GeV. 
In principle, the technology can be extended down to 500~MeV, but these low energies are highly populated with atmospheric backgrounds.
The highest DM mass that we consider in this study is 25~GeV and in this energy range one can expect that 
the muons and electrons produced via CC processes will be fully contained in these massive detectors. 
We assume an 80~\% detection efficiency for electrons and positrons and for $\mu^{\pm}$ we consider a full 100~\%. 
Electron tracks would remain fully contained up to very high energies enabling to probe DM masses up to 100~GeV.
Since magnetization of this detector would be challenging, the
charge of the particle is not measured and we combine the $e^-$ and $e^+$ as well as the 
$\mu^-$ and $\mu^+$ samples.
We assume an energy resolution 
of $\delta E(E)$ = 0.15$\sqrt{E/\rm 1~GeV}$~GeV for CC $\mu^{\pm}$ and $e^{\pm}$ events. 
The atmospheric neutrino 
background can be suppressed to a very low level by selecting a narrow sky window of the size of the angular resolution of the detector centered around the Sun.
The reconstruction of the incoming neutrino direction in a LAr detector has been estimated using the information
coming from all the final state particles and the detector angular resolution as a function of incident neutrino energy
has been shown in~Fig.~\ref{fig:resolution}. This information has been obtained from Fig.~4 of~\cite{Bueno:2004dv}.   
Notice that by using information on both the leptonic and hadronic final states of the interaction, angular resolutions smaller than the typical scattering angle between the neutrino and the charged lepton can be achieved, something impossible if only the outgoing charged lepton is detected. We use the same angular resolution to estimate the atmospheric 
neutrino backgrounds for $\nu_e$/$\bar\nu_e$  and $\nu_{\mu}$/$\bar\nu_{\mu}$ CC signal events.  

\subsection{Magnetized Iron Calorimeter detector}

Iron calorimeters are made up of iron (steel) modules interspersed with sensitive elements in which 
charged particles deposit energy. Electrons and positrons cannot pass through iron and therefore these detectors
are blind to $\nu_e$/$\bar\nu_e$. This type of detectors is well suited to observe the long muon tracks and hence
are capable of observing only  $\nu_{\mu}$ and $\bar\nu_{\mu}$ events. A magnetic field can be added to these
detectors to distinguish $\nu_{\mu}$ events from $\bar\nu_{\mu}$.
A Magnetized Iron Neutrino Detector (MIND)~\cite{Cervera:2000vy,Cervera:2010rz} of large mass (50 - 100~\kton{}) 
has been proposed for the precision study of neutrino oscillations of neutrinos from muon decays at a future neutrino factory facility~\cite{Geer:1997iz,DeRujula:1998hd,Bandyopadhyay:2007kx,ids,Abe:2007bi}. A similar type of detector is being 
considered for the INO facility~\cite{tifr-ino,Agarwalla:2009xc} which is planned to have a total mass of 50~\kton{} at startup, 
which might be later upgraded to 100~\kton{}. The main aim of this detector is to measure atmospheric 
neutrinos~\cite{Raychaudhuri:2008zz} and it may serve as a neutrino factory far detector 
at later stage~\cite{Agarwalla:2010hk}.   

In our study, we consider a 100~\kton{} MIND detector with an energy dependent efficiency as given in \Ref~\cite{ThesisLaing}.
For $\nu_{\mu}$, it remains almost flat ($\sim$ 63\%) in the neutrino energy range of 10 to 25 GeV while for $\bar\nu_{\mu}$,
it deteriorates from 76\% to 71\% going from 10 to 25 GeV. This detector can provide accurate measurements 
of the muon energy~\cite{ThesisLaing}. Moreover, due to the calorimetric 
nature of the detector, the energy of the hadronic shower can also be 
reconstructed~\cite{Michael:2008bc,Adamson:2006xv} which helps to reconstruct the energy of the 
interacting neutrino. We use an energy resolution of $\delta E(E) = 0.15E$ for CC $\mu^{\pm}$ events. 
The angular resolution of this detector is also quite good and mainly governed by the 
hadronic shower~\cite{Bari:2003bt}. We use the information from both muons and hadronic showers~\cite{ThesisLaing,Bari:2003bt}
to reconstruct the incoming neutrino direction. The neutrino angular resolution of this detector 
as a function of true neutrino energy has been depicted in Fig.~\ref{fig:resolution}. The angular resolution for 
$\nu_{\mu}$ is slightly better than $\bar\nu_{\mu}$. We use this information on angular resolution 
to suppress the atmospheric $\nu_{\mu}$ and $\bar\nu_{\mu}$ backgrounds in a similar way to the one discussed above for the 
LArTPC, \ie{} by discarding all events except those reconstructed 
within the solid angle subtended by the neutrino angular resolution centered around the Sun. Notice that the efficiency as well as the energy and angular resolutions of the MIND detector will depend on the number of active layers the track crosses and, therefore, on the direction of the incoming neutrino. While for a neutrino beam experiment the detector can be oriented along this direction this would be more challenging for neutrinos from the Sun. We have neglected this dependence in this first study since it will critically depend on the unknown orientation and latitude of the detector.

\section{Neutrino Fluxes and Event Rates}
\label{sec:fluxesandrates}

If DM particles can pair annihilate we can envision a wide variety of possible channels into lighter SM components. The rates for these annihilations will be greatly enhanced in regions with larger density of DM particles. If the cross section between DM and nuclei is high enough, the collisions of DM with solar matter will allow them to lose energy and become gravitationally bound, accumulating in the solar gravitational potential. Thus, we can expect a larger number of DM annihilations in the Sun than in the surrounding space. Most of the annihilation products will be lost in scatterings and decays in the dense solar interior. Fortunately, neutrinos once again provide us with a unique tool to explore these secretive processes that could take place in the interior of the Sun. Indeed, neutrinos from DM annihilations could escape the solar medium unaffected and provide a picture of the processes taking place in the star's interior. If these neutrinos have energies in excess of GeV they will be clearly distinguishable from the MeV neutrinos produced in solar reactions. 

Many of the possible DM annihilation channels could indeed lead to such neutrino signals~\cite{Cirelli:2005gh}. Assuming that DM annihilations are flavour conserving and predominantly to two body final states, the $b\bar{b}$, $c\bar{c}$, $\tau^-\tau^+$ or $\nu_\alpha\bar{\nu}_{\alpha}$ channels would lead to the required high energy neutrino probes of solar DM annihilations. For high enough DM masses, additional channels leading to high energy neutrino fluxes would also become kinematically allowed such as $t\bar{t}$, $W^+W^-$, $ZZ$ or even Higgs production. The energies of these neutrinos would make them unsuitable for the detector technologies under study and more adequate for large neutrino telescopes, since the associated charged leptons will not be fully contained in a small detector volume and the energy and angular resolution would be greatly degraded. We will therefore limit our study to DM masses up to 25~GeV. Other possible channels such as $e^-e^+$ or $\gamma\gamma$ would not lead to neutrino signals but could provide complementary probes in gamma ray or positron searches.

\begin{figure}
\begin{center}
\includegraphics[width=.48\textwidth]{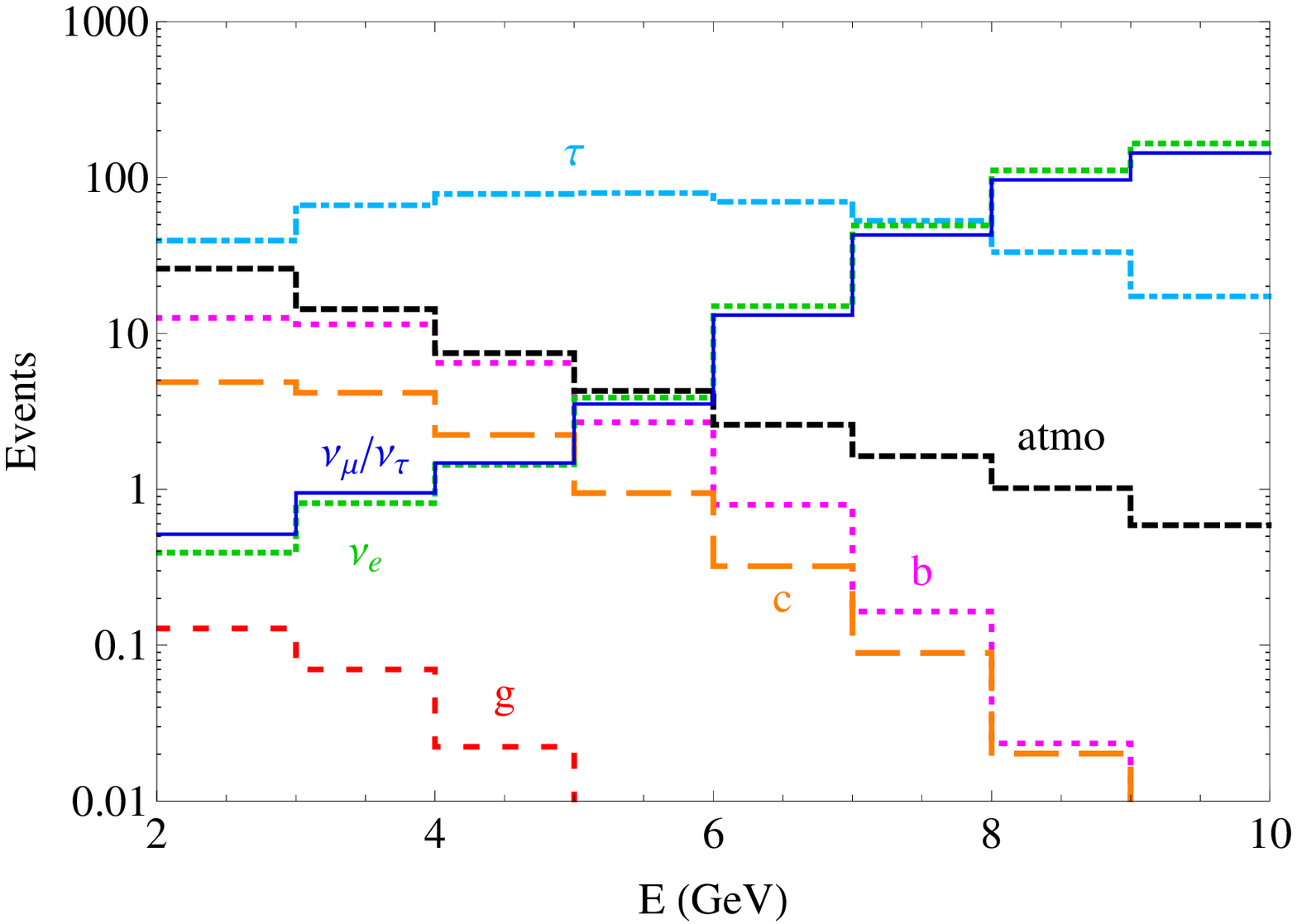}
\includegraphics[width=.48\textwidth]{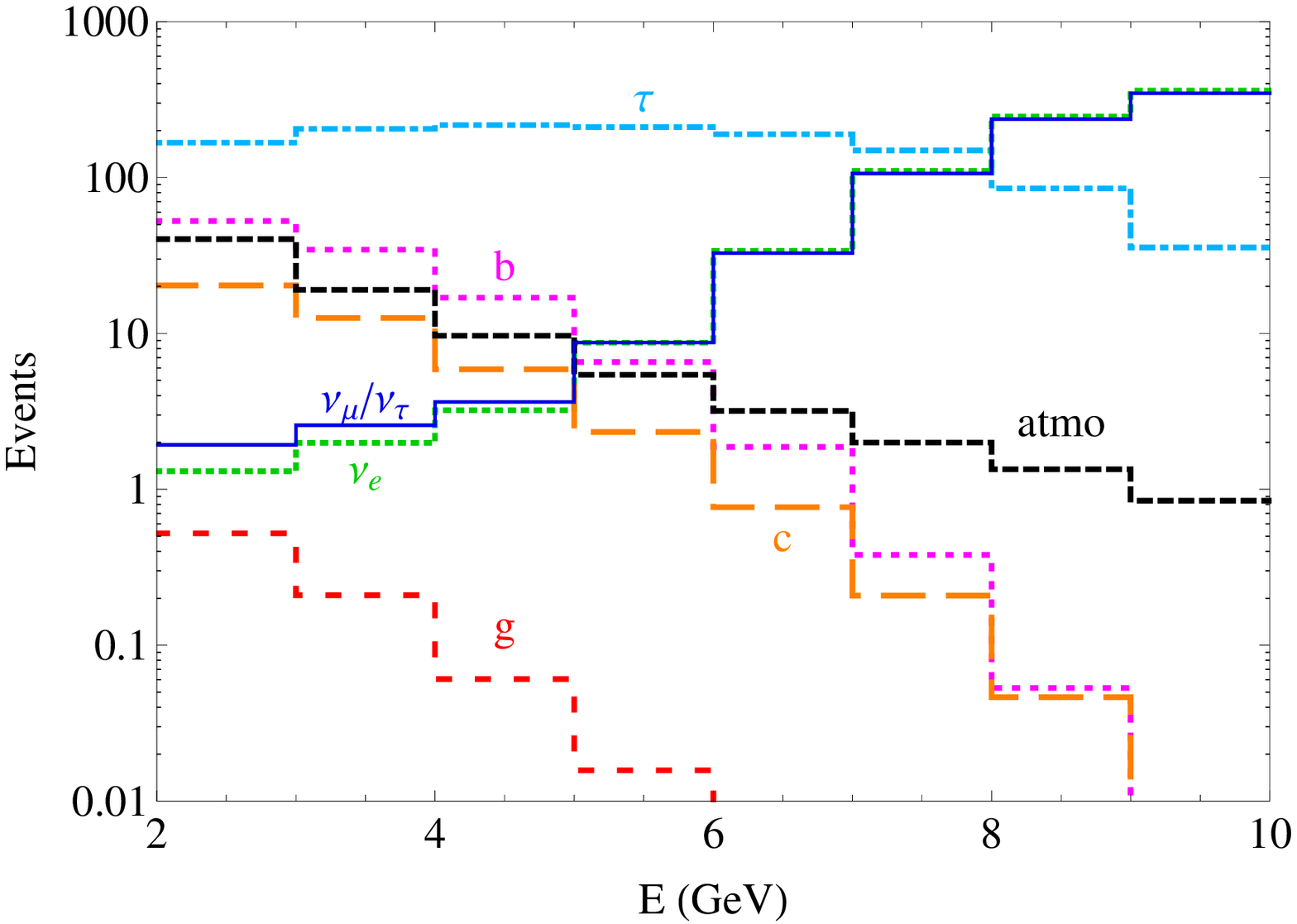} \\
\includegraphics[width=.48\textwidth]{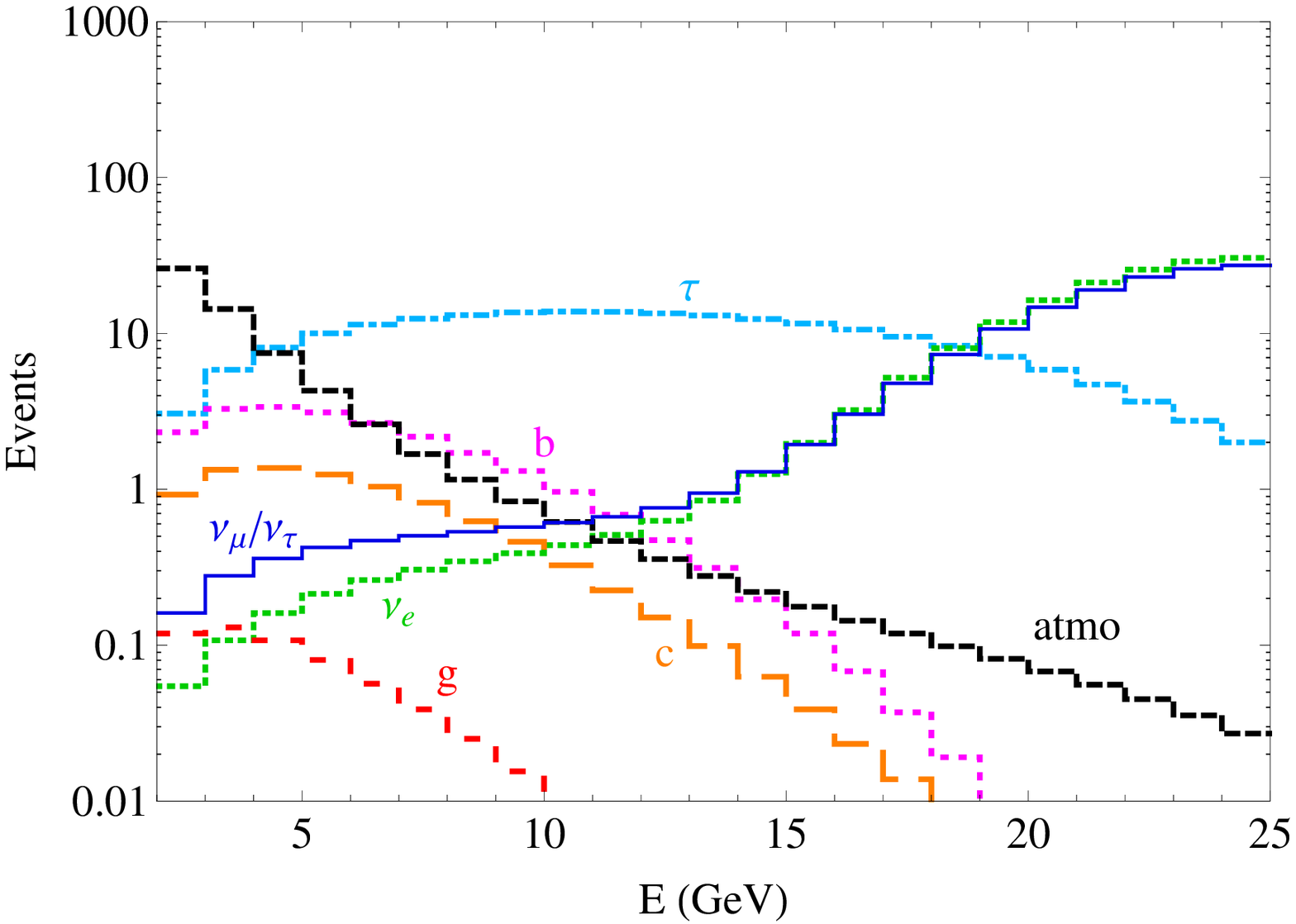}
\includegraphics[width=.48\textwidth]{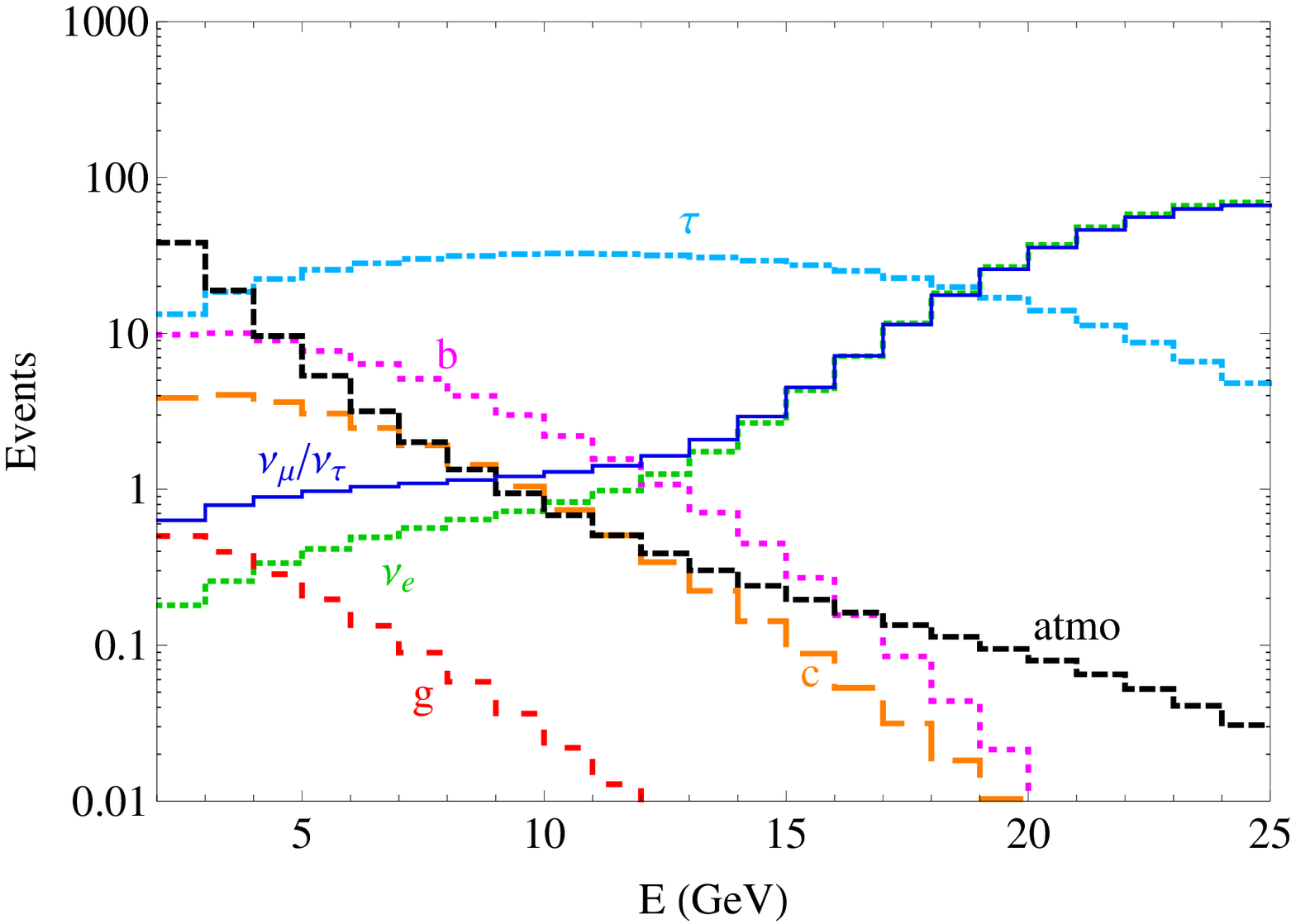}
\caption{Expected muon events in bins of 1~GeV width for the MIND (left) and GLACIER (right) 100~kt detectors and for neutrinos from the annihilation of DM particles of 10~GeV (up) and 25~GeV (down) mass. Ten full years of data taking for a 1~fb DM-nucleon spin-dependent cross section have been assumed. Since GLACIER has no charge identification both $\mu^+$ and $\mu^-$ events where added, while only $\mu^-$ events are displayed for MIND.}
\label{fig:fluxes}
\end{center}
\end{figure}

The expected neutrino fluxes per annihilation at the detector from the different DM annihilation channels have been obtained using the WimpSim~\cite{Blennow:2007tw} software. The fluxes for the ``standard oscillations'' with $\theta_{13}=0^\circ$, $\theta_{23}=45^\circ$, $\theta_{12}=33.2^\circ$, $\Delta m^2_{31}=2.2 \cdot 10^{-3}$ eV$^2$ and $\Delta m^2_{21}=8.1 \cdot 10^{-5}$ eV$^2$ were used. These fluxes must later be normalized with the annihilation rate in the Sun. In equilibrium this annihilation rate is one half of the capture rate, which is proportional to the DM-nucleon cross section and can be computed given the solar density profile, for which we used the standard BP2000 Solar Model~\cite{Bahcall:2000nu}, a local DM density of $\rho_{DM}=0.3$~GeV~cm$^{-3}$, and a Maxwellian velocity distribution shifted by the solar velocity $v_\odot = 220$~km~s$^{-1}$. For the computation of the capture rate we follow~\cite{Gould,Jungman:1995df}. Notice that large systematic astrophysical uncertainties affect the computation of the capture rate and thus the overall normalization of the expected signal. We will not include this systematic error in the numerical results presented in \sec~\ref{sec:reconstruction} since their effect in the sensitivity to the annihilation cross section to the different channels is just a trivial rescaling by the corresponding amount.
These fluxes were later convoluted with the cross section and detector efficiencies and energy resolutions summarized in \tab~\ref{tab:detector} in order to obtain the expected number of event rates depicted in Fig.~\ref{fig:fluxes}.
The event rates have been computed for 10 full years of data taking with a DM flux computed assuming a 1~fb cross section between DM and nucleons. For each channel a 100~\% branching ratio is shown. For the MIND detector the muon neutrino events arising from the different annihilation channels are depicted. Since MIND is a magnetized detector, muon antineutrinos can also be distinguished. The event rates for antineutrinos are very similar in shape but slightly lower than the neutrino ones, so we do not reproduce them again in a separate figure but will include them in the numerical analysis presented in \sec~\ref{sec:reconstruction}. On the other hand, GLACIER is not magnetized and neutrino and antineutrino events need to be summed. However, GLACIER is also suited to detect electron neutrino events. In \fig~\ref{fig:fluxes} we show the muon neutrino and antineutrino events for the GLACIER detector. Electron events have a similar shape and we do not show them but they are also included in the analysis of \sec~\ref{sec:reconstruction}. As expected, the annihilation channels that lead to larger neutrino fluxes at the detector are the neutrino channels, with the hardest spectrum; the $\tau$ channel with an intermediate spectrum; and the heavy quark channels ($b$ and $c$) with a softer spectrum and somewhat smaller flux. The gluon channel ($g$), while also leading to neutrinos, provides an event rate too low to be observed and will be disregarded.

Notice that all the neutrino channels provide a very similar spectrum (since the kinematics of the process is identical) and, moreover, neutrino oscillations ensure that almost equal amounts of all neutrino flavours will arrive at the detector. The $\nu_\mu$ and $\nu_\tau$ channels are indeed indistinguishable (since $\theta_{23}$ was set to maximal mixing), and the $\nu_e$ annihilation channel only shows some deviation at low energies where the event rate is too low to be observed. Thus, when trying to reconstruct the DM annihilation channels in \sec~\ref{sec:reconstruction} we will not distinguish between the different flavours but only consider a generic ``neutrino channel'' for which the $\nu_\mu$ case was chosen. Similarly, the annihilations into $b$ and $c$ quarks provide very similar spectra. As will be shown with a particular example in \sec~\ref{sec:reconstruction}, the detector energy resolution is not enough to discriminate between these two channels and an effective ``quark channel'' $q$ will be considered instead, for which the $b$ case was chosen as representative.   

A common feature of all the channels is that the corresponding event rates will have an abrupt cutoff at the energy corresponding to the DM mass, providing a clean and unambiguous measurement of this parameter. This will be true as long as either the $\tau$ or neutrino signal is measured, given their harder spectrum. Thus, in the analysis of \sec~\ref{sec:reconstruction} the DM mass will be fixed, since it is not degenerate with any other parameter. The free parameters that we will try to reconstruct will be the product of the DM-nucleon cross section times the branching ratio of DM annihilations to neutrinos, $\tau$ and heavy quarks.   

\subsection{Atmospheric Neutrino Background}
\label{sec:seciia}

The main source of background for the search of neutrinos from DM annihilations in the Sun will be the atmospheric neutrino events at the detector. Indeed, unlike solar neutrinos, the atmospheric neutrino energy range overlaps with that of the signal. This is an unavoidable and potentially large background which can fortunately be reduced by exploiting the extremely good angular and energy resolution of the detectors under consideration. Indeed, while the signal neutrinos come from the direction of the Sun, the atmospheric neutrinos reach the detector from all directions and can be significantly reduced by accepting only the background events for which the reconstructed incoming neutrino direction corresponds with the solar position at the time 
of the event.

In order to simulate the atmospheric neutrino background we take as representative fluxes those from \Ref~\cite{Honda:2011nf} for the Frejus site with mountain over detector at a solar maximum. These fluxes are then oscillated with the same oscillation parameters as the DM signal and convoluted with the cross section, detector efficiency and energy resolutions summarized in \tab~\ref{tab:detector}. In order to estimate the background suppression achieved by the angular cut around the solar position, we first estimate the number of atmospheric events for a given energy integrating over the 
whole sky and then multiply the number of events by the solid angle subtended by a cone of the detector angular resolution at the corresponding energy divided by $4 \pi$. Of course this is only an approximation and the atmospheric flux should only be integrated over the position of the Sun at a particular time. However, since this will depend on what is the actual geographical site selected for the detector, we regard this average over all the sky as a good approximation of the magnitude of the atmospheric neutrino background. We have considered only CC events in estimating the atmospheric neutrino background and in the numerical analysis we add a 
10~\% systematic uncertainty (fully correlated in all the energy bins) to this background. The contribution from 
neutral current (NC) events is negligible for both electron and muon samples due to the high background rejection capability  ($>$~99.5~\%) of the detectors under consideration. The atmospheric $\nu_{\tau}$/$\bar\nu_{\tau}$ fluxes produced via neutrino oscillations can give rise to tau leptons through CC interactions inside the detector. 
These taus can decay further into electrons and muons with a branching ratio of~$\sim$~17~\%. We have checked that these backgrounds are negligible. 

The black dashed lines in Fig.~\ref{fig:fluxes} depict the estimated atmospheric neutrino background for each detector after the angular cut. As can be seen, the background decreases fast with the neutrino energy and mainly affects the observability of the quark annihilation channels. Indeed, we have checked that increasing the background mainly affects the sensitivity to this channel, while the neutrino and $\tau$ ones remain essentially unaffected.

\section{Reconstructing the Dark Matter Annihilation Branching Ratios}
\label{sec:reconstruction} 

As motivated in the previous section, we will scan and analyze the parameter space consisting of ${\rm BR}_x \sigma$ ($x = \tau,\nu,q$) in order to determine the capabilities of the two detectors under study to discriminate between the different DM annihilation channels. This would provide invaluable information to shed light on the DM nature beyond the mass and cross section measurement that is available at direct searches and complementary to the other annihilation channels explored in indirect searches with gamma-ray satellites and telescopes. We will also explore the sensitivities that both detectors have to these parameters. The parameter space has been sampled using a MCMC based on the MonteCUBES~\cite{Blennow:2009pk} software. We perform this scan for simulated true values of both $\sigma = 0$ and ${\rm BR}_x \sigma = 0.3$~fb. The running time of the experiments were set to 10~years each with the efficiencies and energy resolutions summarized in Table~\ref{tab:detector}. For the MIND detector we separate the signal in muon neutrinos and antineutrinos, exploiting the magnetization of the detector to separate the two samples. For the GLACIER detector magnetization would be challenging and we therefore combine the neutrino and antineutrino samples. On the other hand, we exploit the capability of GLACIER of discerning the neutrino flavour and consider electron and muon neutrinos separately. In all cases, the threshold was set to 2~GeV and the events were distributed in bins of 1~GeV width.

Since we are simultaneously scanning over all the parameters, we can easily see whether or not there are any correlations among the different branching ratios and what are the actual sensitivities to the parameters when marginalizing over the others. Regarding the DM mass, which could a priori present degeneracies with the other parameters under study, its determination is expected to be free of degeneracies with other parameters since it corresponds to the cutoff of the neutrino spectra and will be easily detected for annihilations into neutrinos or taus. Thus, we will keep this parameter fixed in the present study for simplicity. As we will show, in general, the sensitivities to the different branching ratios turn out to be fairly independent, since the sum will be bounded from above by the total event rate.

For every scan in the MCMC, we generate four chains with a total number of $2\cdot 10^{5}$ samples per chain, check the variance within each chain and compare it with the variance of the overall sample in order to gauge the convergence of the chains. In all cases presented, we obtained a good convergence ($R-1 < 10^{-3}$)\footnote{See \Ref~\cite{Gelman:1992zz} for the definition of this measure.}. The MCMC priors were assumed to be flat in the scanned parameters.

\subsection{Results}
\label{sec:results}

\begin{figure}
\centering
\includegraphics[width=0.5\textwidth]{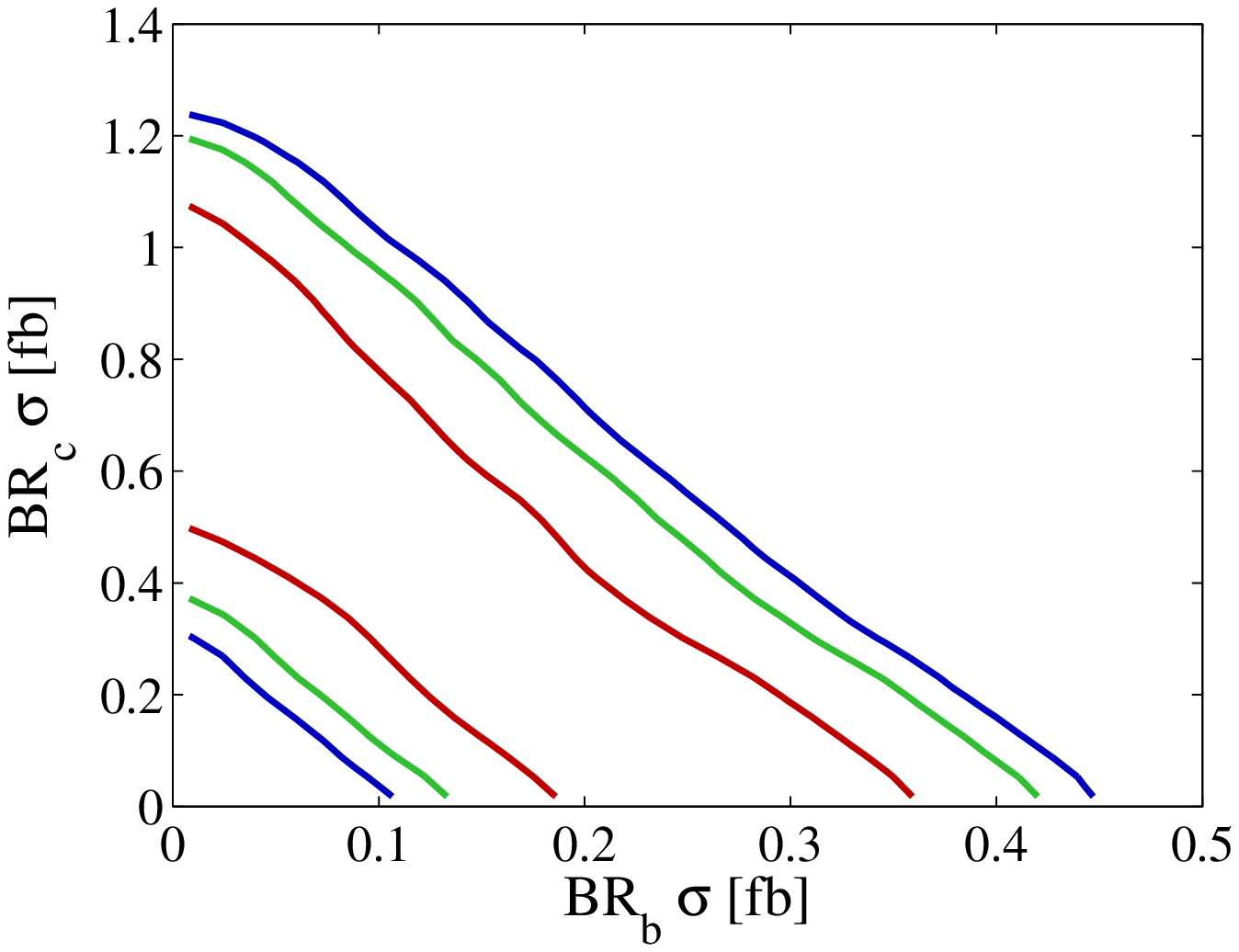}
\mycaption{The most probable regions at 68~\%, 90~\%, and 95~\% posterior probability and a DM mass of 10~GeV projected onto the $\BR_b\sigma-\BR_c\sigma$ plane. This simulation was made for GLACIER with a DM mass assuming a true $\BR_b\sigma = 0.3$~fb as the only source of neutrinos in DM annihilations. We can clearly see the linear degeneracy between the two parameters, motivating the study of only an effective quark channel $\BR_q\sigma$.\label{fig:degeneracy}}
\end{figure}
In \fig~\ref{fig:degeneracy}, we illustrate the degeneracy between the DM annihilations into $b$ and $c$ quarks with a simulation for GLACIER with a DM mass of 10~GeV and $\BR_b \sigma = 0.3$~fb and keeping $\BR_b \sigma$ and $\BR_c \sigma$ as free parameters in the fit. As we will see later, this corresponds to the most favourable setting and, from this figure, it is apparent that the two channels cannot be distinguished, and thus it makes sense only to fit the combination $\BR_q = \BR_b + k \BR_c$ in our analysis, where $k \simeq 0.4$ accounts for the smaller number of neutrinos resulting from each annihilation into $c\bar c$. In order to simulate this channel we take the events expected from DM annihilation into $b\bar b$.

In \tab~\ref{tab:sensitivity}, we present the results for our MCMC simulations for the case when there is no signal. The numbers correspond to the 90~\% limit on the posterior probability and can be interpreted as the sensitivity reach of the experiment to each parameter. From the table, we can clearly see that the GLACIER detector would be significantly more sensitive to DM annihilations at low DM mass than the MIND one, having typical sensitivity of a few $10^{-2}$~fb for both ${\BR}_\tau \sigma$ and ${\BR}_\nu \sigma$, where the corresponding MIND sensitivies are $\mathcal O(0.1-1)$~fb.
There are several reasons why the MIND sensitivities are worse than the ones for GLACIER. First of all, the signal statistics are higher at GLACIER, since it considers both electrons and muons, and because of a higher assumed efficiency. In particular, the MIND efficiency is very bad at low energies, leading to the significantly worse bounds for the case of 10~GeV DM mass. Furthermore, the angular and energy resolutions of the LArTPC are better, leading to a better background rejection in these experiments.

These sensitivity limits should be compared to the current bounds on the spin-dependent DM cross section. The most stringent direct bounds at the simulated masses are $\sigma \lesssim 400$~fb at 10~GeV (PICASSO~\cite{Archambault:2009sm}) and $\sigma \lesssim 90$~fb at 25~GeV (COUPP~\cite{Behnke:2010xt}). For the 10~GeV case, the present exclusion limit of PICASSO is also close to the lower part of the region required to explain the DAMA/LIBRA signal when interpreted as spin-dependent scatterings. Thus, depending on the dominating annihilation channel, the current bounds could be improved by up to five orders of magnitude in the most optimistic case of GLACIER, a DM mass of 10~GeV, and annihilations directly into neutrinos. However, it should be pointed out that DM is not required to have any branching ratio at all into channels that produce a neutrino flux, and thus direct detection experiments will always be complementary to indirect searches. However, if an annihilation branching ratio of at least $\sim 10^{-3}$ to heavy quarks, $\sim 10^{-4}$ to $\tau^+ \tau^-$ or $\sim 10^{-5}$ to neutrinos is present, the DM spin-dependent scattering interpretation of the DAMA/LIBRA signal can be probed at the GLACIER detector. Similarly, for a spin-independent scattering interpretation of DAMA/LIBRA, branching ratios of $\sim 10^{-1}$ to quarks, $\sim 10^{-2}$ to $\tau^+ \tau^-$ and $\sim 10^{-3}$ at GLACIER could be probed.

In comparison to the results obtained for Super-K~\cite{Kappl:2011kz}, we find that the MIND experiment would be of comparable sensitivity, while any of the LArTPC experiments studied would out-perform Super-K in most cases.
However, it has to be borne in mind that the capabilities of a liquid argon detector have been significantly less studied and simulated than for magnetized iron calorimeters such as MIND, thus, the better performance of GLACIER is conditioned to achieving the optimistic assumptions listed in \tab~\ref{tab:detector}. Typically, the sensitivity to ${\BR}_q \sigma$ is significantly worse due to the softer spectrum of this channel that makes it more contaminated by the atmospheric neutrino background. Indeed, we have verified that increasing the atmospheric background with less restrictive cuts greatly deteriorates the reach in ${\BR}_q \sigma$ while having a small impact in the measurement of the other two parameters. Moreover, as can be seen from \fig~\ref{fig:fluxes}, a smaller flux of neutrinos per annihilation as compared to the other annihilation channels is obtained for the quarks. Conversely, the present bounds on this channel are also weaker and a MIND detector would actually provide some improvement on this channel as compared to the Super-K bounds given its better angular resolution and, hence, background rejection. It is also interesting to note that the sensitivity of the GLACIER detector decreases with the DM mass assumed, as expected from the reduced neutrino flux (\cf~\fig~\ref{fig:fluxes}). On the other hand, the sensitivity of the MIND detector increases with the DM mass demonstrating the better optimization of this detector to high neutrino energies.  
\begin{table}
\begin{center}
\begin{tabular}{|ll|ccc|}
\hline
Experiment & DM mass & ${\BR}_{\tau} \sigma$~[fb] & ${\BR}_{\nu} \sigma$~[fb] & ${\BR}_q \sigma$~[fb]  \\
\hline
MIND (100~\kton{})            & 10~GeV    & 0.70                & 0.35               & 3.4                \\
                 & 25~GeV    & 0.34                & 0.15               & 1.7                \\
\hline
LArTPC (34~\kton{})   & 10~GeV    & 0.15                & 0.11               & 0.73               \\
                 & 25~GeV    & 0.16                & 0.10               & 0.21               \\
\hline
GLACIER (100~\kton{})         & 10~GeV    & $1.5\cdot 10^{-2}$  & $6.4\cdot 10^{-3}$ & 0.25               \\
                 & 25~GeV    & $1.0\cdot 10^{-2}$  & $5.2\cdot 10^{-3}$ & 0.19               \\
\hline
Super-K data \cite{Kappl:2011kz} & 10~GeV & 0.65        & 0.12               & 10                 \\
                 & 25~GeV    & 0.45                & 0.19               & 5.0                \\
\hline
\end{tabular}
\mycaption{Sensitivities to the branching ratios multiplied by the capture cross section at 90~\%
posterior probability for different experiments and DM mass after 10 years data taking. The Super-K sensitivities are taken from \fig~4 of \Ref~\cite{Kappl:2011kz}.}
\label{tab:sensitivity}
\end{center}
\end{table}

\begin{figure}
\centering
\includegraphics[width=0.32\textwidth]{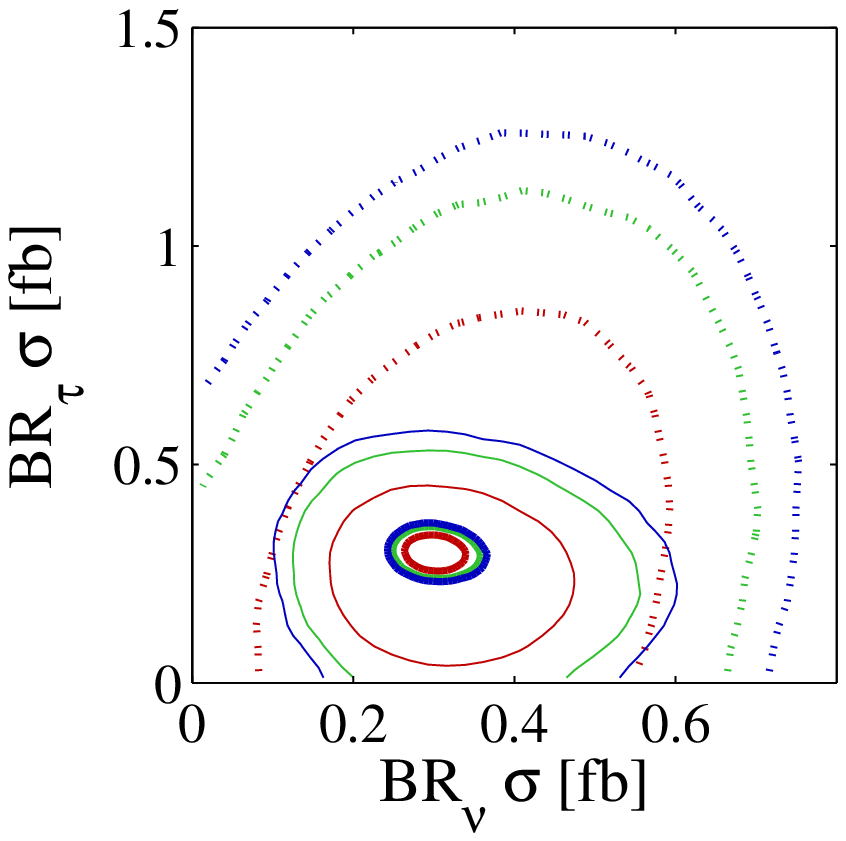}
\includegraphics[width=0.32\textwidth]{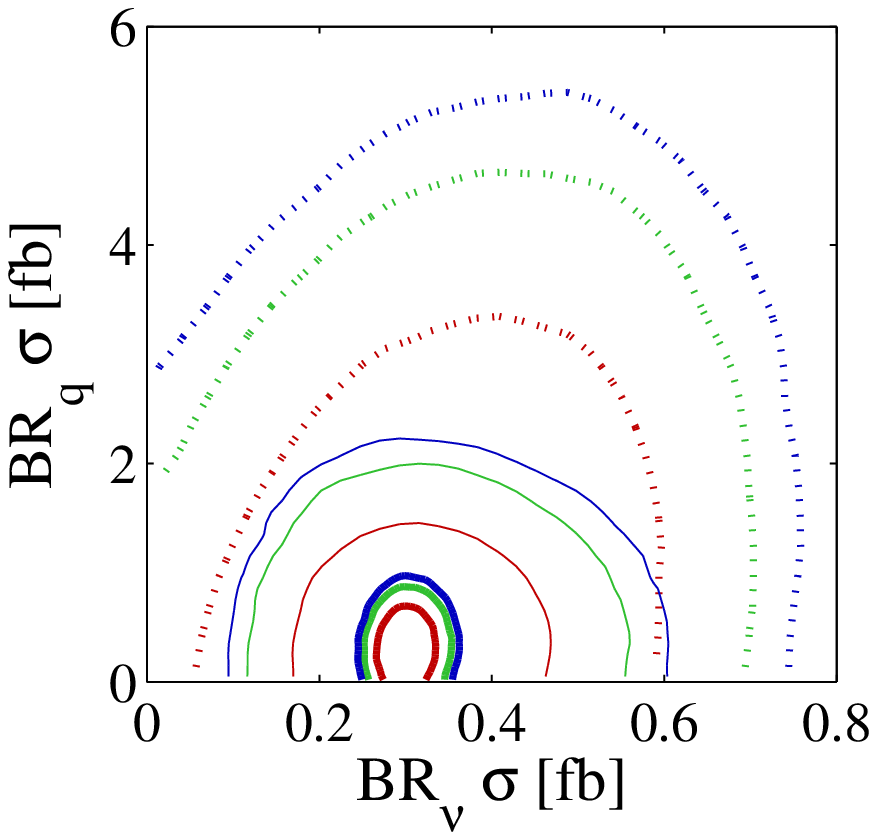}
\includegraphics[width=0.32\textwidth]{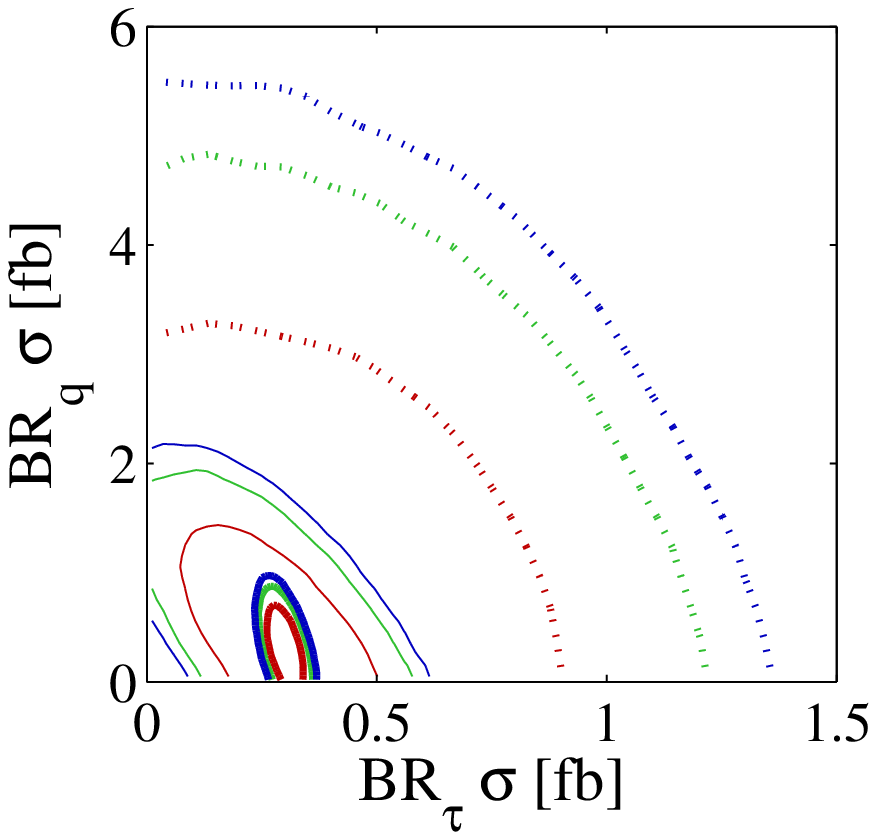} \\
\includegraphics[width=0.32\textwidth]{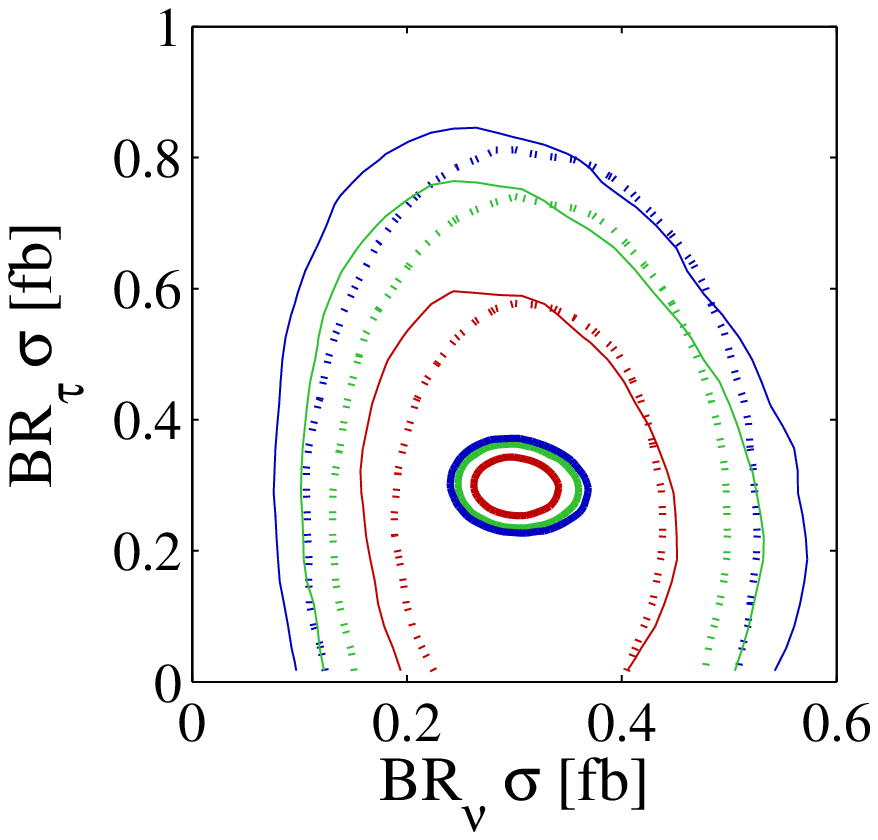}
\includegraphics[width=0.32\textwidth]{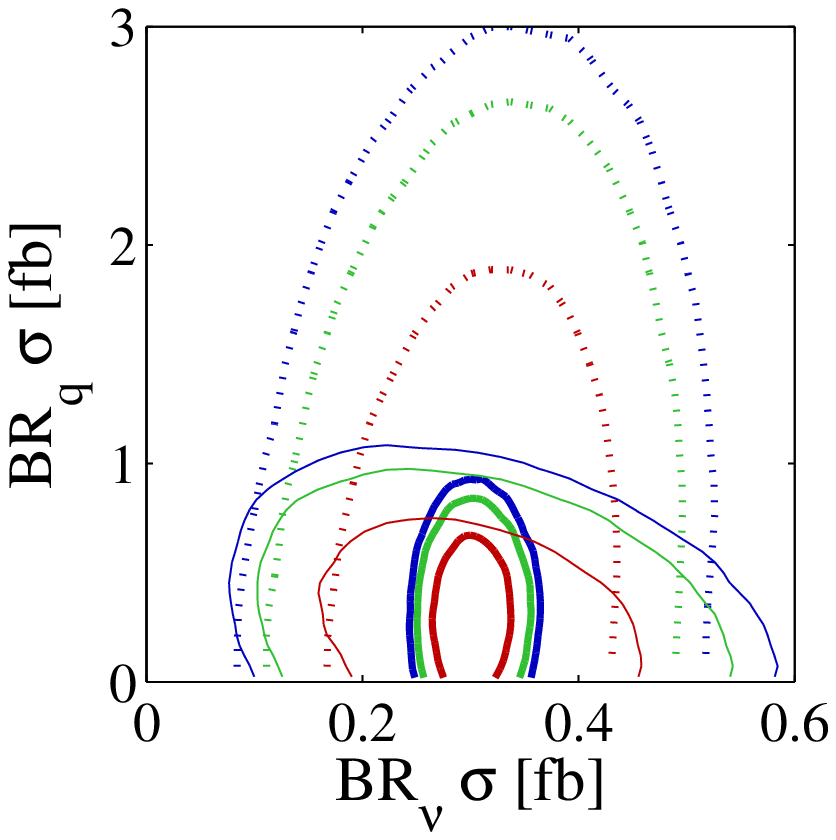}
\includegraphics[width=0.32\textwidth]{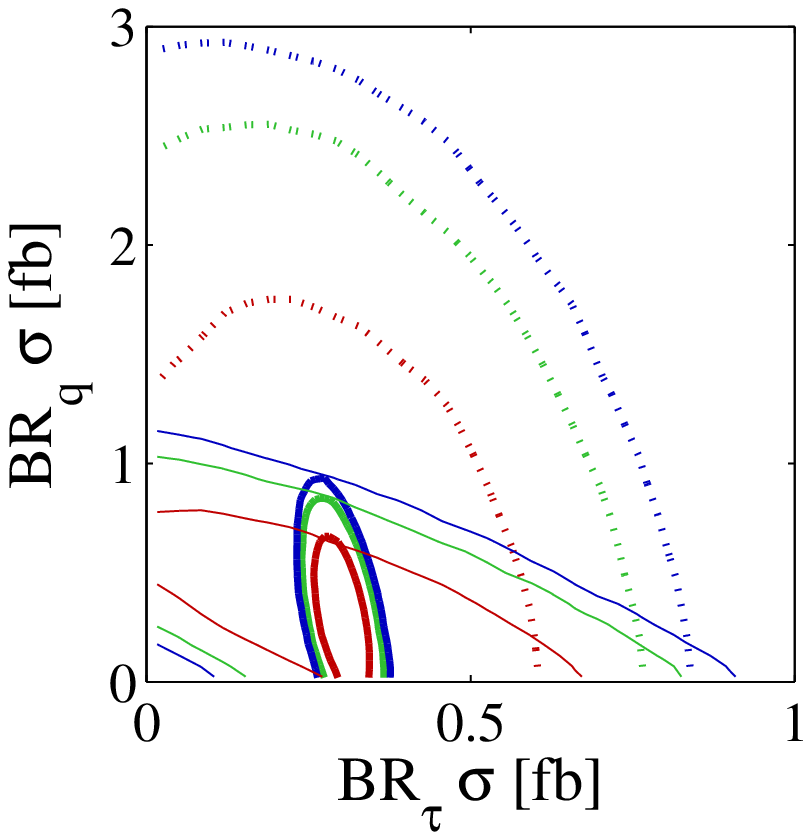}
\mycaption{Typical results for the most probable regions of ${\rm BR}_x \sigma$ ($x = \tau,\nu,q$) at 68~\%, 90~\%, and 95~\% posterior probability and a DM mass of 10 (upper pannels) and 25~GeV (lower panels). The solid lines show the GLACIER (thick) and 34~\kton{} LArTPC (thin) results, while the dotted lines show the MIND results. The results correspond to 10~years of exposure for GLACIER, the 34~\kton{} LArTPC and MIND.\label{fig:10measure}}
\end{figure}

In \fig~\ref{fig:10measure} we show the preferred regions of the parameter space in the case of simulated $\BR_x \sigma = 0.3$~fb. As expected from the sensitivities, the MIND detector is only capable of distinguishing $\BR_\nu \sigma$ from zero, while the signal for $\BR_\tau \sigma$ is close to being detectable. On the other hand, GLACIER would be able to clearly identify the nature of the DM branching ratios into both $\tau$s and $\nu$s, and being close to the sensitivity needed for detecting $\BR_q\sigma$ at this level.
The reproduction of the parameters by GLACIER is remarkable and essentially free of correlations between the different branching ratios. There is a very small correlation between $\BR_\tau$ and $\BR_q$, since both of these spectra are both softer than the $\BR_\nu$ one (\cf~\fig~\ref{fig:fluxes}), but this correlation is almost negligible. Since the different branching ratios of DM could be singled out in an experiment like this, we would gain valuable information on the DM couplings to different particles, which in turn would be of large interest to constrain different models of DM. This would not be true in, \eg, direct detection experiments, which would only be sensitive to the total cross section on the target. It is important to remark that the discerning capability of the detectors between the different DM annihilation channels stems almost exclusively from their energy resolution and the ability to reconstruct the different spectral shapes shown in \fig~\ref{fig:fluxes}. On the other hand, the capability of the MIND detector to separate the neutrino and antineutrino samples does not help with this measurement since the corresponding antineutrino event shapes are extremely similar. Similarly, the information on the neutrino flavour GLACIER provides is largely irrelevant for the analysis and its ability to detect $\nu_e$ in addition to $\nu_\mu$ only provides an increase in statistics but no complementary information to solve degeneracies. The slight degeneracy between $\BR_\nu \sigma$ and the other parameters, in particular for the 10~GeV DM mass results for MIND, is mainly due to the 10~\% systematic error assumed for the atmospheric background simulation and the energy resolution and statistics not being sufficient to tell the spectral shape of the background from a combination of $\BR_\nu \sigma$ and $\BR_q \sigma$. Notice that this effect also worsens the sensitivity estimated for the MIND detector, since relatively large values of $\BR_\nu \sigma$ and $\BR_q \sigma$ simultaneously cannot be disentangled from a situation with a smaller atmospheric background accommodated within the 10~\% systematic allowed. Thus, the sensitivity limits derived here, in particular for the MIND detector, are more conservative than the ones presented for Super-K in \Ref~\cite{Kappl:2011kz} since only one branching ratio a time was considered there and in  order for the signal to mimic the background a combination of both $\BR_\nu \sigma$ and a softer channel is necessary, as depicted by the degeneracy in the top left and top center panels of \fig~\ref{fig:10measure}.

\section{Summary}
\label{sec:summary}

In this paper we have studied the detection of neutrinos as a possible indirect signal of low mass (10--25~GeV) DM annihilating in the Sun
at the next generation of neutrino detectors, as well as the determination of which annihilation channels are present.
Due to the results of the DAMA/LIBRA, CoGeNT and CRESST collaborations, this mass range, which is unaccessible to neutrino telescopes due to the low neutrino energy, is of great interest.
We have focused mainly on the LArTPC and MIND detector technologies. Apart from having good energy and angular resolution capabilities, these two detector types are under consideration for several future neutrino experiments. For the LArTPC, we consider both a 34~\kton{} and a 100~\kton{} version, corresponding to detectors proposed for the LBNE and LAGUNA projects, respectively, and for the MIND we will consider a 100~\kton{} detector, corresponding to the neutrino factory proposal or an upgraded version of INO.

For the signal rate, we computed the solar capture rate of DM following \Ref~\cite{Gould,Jungman:1995df} and using the standard BP2000 Solar Model~\cite{Bahcall:2000nu}. Our results are presented in the variables $\BR_{x}\sigma$ ($x = \tau,\nu,q$), where $\BR_x$ is the annihilation branching ratio into $x\bar{x}$ and $\sigma$ is the total spin-dependent DM-nucleon cross section. We have not included a systematic error on the signal, since this will mainly be an overall factor due to different capture mechanics and therefore lead to a simple rescaling of the $\BR_x \sigma$ parameters.

The main background to our signal is atmospheric neutrinos, for which we assume a systematic overall normalization error of 10~\%. In order to suppress this background, we use the fact that the detectors under study have the capability of measuring the energy and direction of the incoming neutrino and demand that this direction is within the detector angular resolution from the direction of the Sun.

Our results show that with 10 years data the facilities under study would have a sensitivity to $\BR_x \sigma$ of $\mathcal O(10^{-2}-1)$~fb (\cf~\tab~\ref{tab:sensitivity}) with the best sensitivity being achieved for the 100~\kton{} LArTPC detector and the worst for the MIND technology. This should be put into context by the current bounds from direct detection experiments, which give $\sigma \lesssim 400$~fb at 10~GeV mass (PICASSO~\cite{Archambault:2009sm}) and $\sigma \lesssim 90$~fb at 25~GeV (COUPP~\cite{Behnke:2010xt}) and the region favoured by the DAMA/LIBRA signal around 10~GeV mass and $\sigma \sim 1$~pb. The latest results by SIMPLE~\cite{Felizardo:2011uw} improve on these limits down to $\sim 10$~fb. The sensitivities to different annihilation channels are essentially independent without correlations among them. For comparison with current experiments, the MIND detector would perform essentially on the level of the current Super-K~\cite{Kappl:2011kz} data, although the analysis presented here is more conservative, allowing for degeneracies among the different channels and the atmospheric background that were not taken into account by the analysis of Super-K data and for which they would be potentially more dangerous, since only a counting experiment with no energy information was assumed in the study. On the other hand, the 100~\kton{} LArTPC would out-perform Super-K by almost two orders of magnitude. In general, the sensitivities are better for the $\nu\bar\nu$ channel and worse for the $q\bar{q}$ channel, due to the spectra being harder and softer, respectively, as well as a lower event rate per annihilation for the $q\bar{q}$ channel, and thus leading to different levels of confusion with the soft atmospheric spectrum. 

We have also simulated the results for a typical signal when $\BR_x \sigma = 0.3$~fb for all $x$. As expected from \tab~\ref{tab:sensitivity}, it would be challenging for all of the considered experiments to distinguish $\BR_q \sigma$ from zero, while the LArTPC detectors would clearly measure $\BR_\nu \sigma$ and to some extent $\BR_\tau \sigma$ (\cf~\fig~\ref{fig:10measure}). We point out that the experiments considered have the remarkable ability to tell apart the different annihilation branching ratios exploiting their energy resolution, something impossible with a counting experiment such as the study performed for present Super-K data. This demonstrates the complementarity of DM searches at neutrino detectors compared to other DM probes in the study of the nature of DM.  

We conclude that the prospects for improving the limits on low mass DM annihilations in the Sun in future neutrino experiments are in general good. Although current results for Super-K rule out the DAMA and CoGeNT signals with large branching ratios into channels giving rise to neutrinos, solutions with subdominant annihilation into these channels are still viable and present bounds can be improved by almost two orders of magnitude at a detector with the characteristics of GLACIER. Furthermore, these future experiments would have some capabilities of measuring the actual branching ratio into different channels by using the shape of the energy spectrum and such a measurement would provide important information on the nature of the DM.


\begin{acknowledgments}
We would like to thank A.~Cervera, J.M.~Conrad, P.~Hernandez, P.~Huber, S.~Palomares-Ruiz, and M.H.~Shaevitz for useful discussions. The work of M.B.\ is supported by the European Union through the European Commission Marie Curie Actions Framework Programme~7 Intra-European Fellowship: Neutrino Evolution, O.M.\ is supported 
by AYA2008-03531 and CSD2007-00060. We also acknowledge the support from the 
European Union under the European Commission Framework Programme~7 Design Study 
EUROnu, Project 212372 and the project Consolider-Ingenio CUP.
\end{acknowledgments}


\begin{thebibliography}{10}

\bibitem{Jungman:1995df}
G.~Jungman, M.~Kamionkowski, and K.~Griest, {\it {Supersymmetric dark matter}},
   {\em Phys. Rept.} {\bf 267} (1996) 195--373,
  [\href{http://xxx.lanl.gov/abs/hep-ph/9506380}{{\tt hep-ph/9506380}}].

\bibitem{Servant:2002aq}
G.~Servant and T.~M. Tait, {\it {Is the lightest Kaluza-Klein particle a viable
  dark matter candidate?}},  {\em Nucl.Phys.} {\bf B650} (2003) 391--419,
  [\href{http://xxx.lanl.gov/abs/hep-ph/0206071}{{\tt hep-ph/0206071}}].

\bibitem{Cheng:2002ej}
H.-C. Cheng, J.~L. Feng, and K.~T. Matchev, {\it {Kaluza-Klein dark matter}},
  {\em Phys.Rev.Lett.} {\bf 89} (2002) 211301,
  [\href{http://xxx.lanl.gov/abs/hep-ph/0207125}{{\tt hep-ph/0207125}}].

\bibitem{Bertone:2004pz}
G.~Bertone, D.~Hooper, and J.~Silk, {\it {Particle dark matter: Evidence,
  candidates and constraints}},  {\em Phys.Rept.} {\bf 405} (2005) 279--390,
  [\href{http://xxx.lanl.gov/abs/hep-ph/0404175}{{\tt hep-ph/0404175}}].

\bibitem{Kappl:2011kz}
R.~Kappl and M.~W. Winkler, {\it {New Limits on Dark Matter from
  Super-Kamiokande}},  \href{http://xxx.lanl.gov/abs/1104.0679}{{\tt
  arXiv:1104.0679}}.

\bibitem{Bernabei:2010mq}
R.~Bernabei {\em et.~al.}, {\it {New results from DAMA/LIBRA}},  {\em Eur.
  Phys. J.} {\bf C67} (2010) 39--49,
  [\href{http://xxx.lanl.gov/abs/1002.1028}{{\tt arXiv:1002.1028}}].

\bibitem{Aprile:2011hi}
{\bf XENON100} Collaboration, E.~Aprile {\em et.~al.}, {\it {Dark Matter
  Results from 100 Live Days of XENON100 Data}},
  \href{http://xxx.lanl.gov/abs/1104.2549}{{\tt arXiv:1104.2549}}.

\bibitem{Angle:2011th}
{\bf XENON10} Collaboration, J.~Angle {\em et.~al.}, {\it {A search for light
  dark matter in XENON10 data}},  \href{http://xxx.lanl.gov/abs/1104.3088}{{\tt
  arXiv:1104.3088}}.

\bibitem{Ahmed:2010wy}
{\bf CDMS-II} Collaboration, Z.~Ahmed {\em et.~al.}, {\it {Results from a
  Low-Energy Analysis of the CDMS II Germanium Data}},  {\em Phys. Rev. Lett.}
  {\bf 106} (2011) 131302, [\href{http://xxx.lanl.gov/abs/1011.2482}{{\tt
  arXiv:1011.2482}}].

\bibitem{Aalseth:2010vx}
{\bf CoGeNT} Collaboration, C.~E. Aalseth {\em et.~al.}, {\it {Results from a
  Search for Light-Mass Dark Matter with a P-type Point Contact Germanium
  Detector}},  {\em Phys. Rev. Lett.} {\bf 106} (2011) 131301,
  [\href{http://xxx.lanl.gov/abs/1002.4703}{{\tt arXiv:1002.4703}}].

\bibitem{Hooper:2010uy}
D.~Hooper, J.~I. Collar, J.~Hall, and D.~McKinsey, {\it {A Consistent Dark
  Matter Interpretation For CoGeNT and DAMA/LIBRA}},  {\em Phys. Rev.} {\bf
  D82} (2010) 123509, [\href{http://xxx.lanl.gov/abs/1007.1005}{{\tt
  arXiv:1007.1005}}].

\bibitem{Schwetz:2010gv}
T.~Schwetz, {\it {Direct detection data and possible hints for low-mass
  WIMPs}},  \href{http://xxx.lanl.gov/abs/1011.5432}{{\tt arXiv:1011.5432}}.

\bibitem{cogent}
``2011 may symposium on dark matter, space telescope science institute
  baltimore, maryland.''
\newblock {\tt https://webcast.stsci.edu/webcast/detail.xhtml?talkid=2426}.

\bibitem{Jochum:2011zz}
J.~Jochum {\em et.~al.}, {\it {The CRESST dark matter search}},  {\em Prog.
  Part. Nucl. Phys.} {\bf 66} (2011) 202--207.

\bibitem{Archambault:2009sm}
S.~Archambault {\em et.~al.}, {\it {Dark Matter Spin-Dependent Limits for WIMP
  Interactions on 19-F by PICASSO}},  {\em Phys. Lett.} {\bf B682} (2009)
  185--192, [\href{http://xxx.lanl.gov/abs/0907.0307}{{\tt arXiv:0907.0307}}].

\bibitem{Behnke:2010xt}
E.~Behnke {\em et.~al.}, {\it {Improved Limits on Spin-Dependent WIMP-Proton
  Interactions from a Two Liter CF$_3$I Bubble Chamber}},  {\em Phys. Rev.
  Lett.} {\bf 106} (2011) 021303,
  [\href{http://xxx.lanl.gov/abs/1008.3518}{{\tt arXiv:1008.3518}}].
  

\bibitem{Felizardo:2011uw}
E.~Behnke {\em et.~al.}, {\it {Final Analysis and Results of the Phase II SIMPLE Dark Matter Search}},{{\tt arXiv:1106.3014}}].

\bibitem{Silk:1985ax}
J.~Silk, K.~A. Olive, and M.~Srednicki, {\it {The photino, the sun, and
  high-energy neutrinos}},  {\em Phys. Rev. Lett.} {\bf 55} (1985) 257--259.

\bibitem{Krauss:1985ks}
L.~M. Krauss, K.~Freese, W.~Press, and D.~Spergel, {\it {Cold dark matter
  candidates and the solar neutrino problem}},  {\em Astrophys. J.} {\bf 299}
  (1985) 1001.

\bibitem{Freese:1985qw}
K.~Freese, {\it {Can Scalar Neutrinos Or Massive Dirac Neutrinos Be the Missing
  Mass?}},  {\em Phys. Lett.} {\bf B167} (1986) 295.

\bibitem{Krauss:1985aaa}
L.~M. Krauss, M.~Srednicki, and F.~Wilczek, {\it {Solar System Constraints and
  Signatures for Dark Matter Candidates}},  {\em Phys. Rev.} {\bf D33} (1986)
  2079--2083.

\bibitem{Desai:2004pq}
{\bf Super-Kamiokande} Collaboration, S.~Desai {\em et.~al.}, {\it {Search for
  dark matter WIMPs using upward through-going muons in Super-Kamiokande}},
  {\em Phys. Rev.} {\bf D70} (2004) 083523,
  [\href{http://xxx.lanl.gov/abs/hep-ex/0404025}{{\tt hep-ex/0404025}}].
  
\bibitem{Niro:2009mw}
V.~Niro, A.~Bottino, N.~Fornengo, and S.~Scopel, {\it {Investigating light
  neutralinos at neutrino telescopes}},  {\em Phys.Rev.} {\bf D80} (2009)
  095019, [\href{http://xxx.lanl.gov/abs/0909.2348}{{\tt arXiv:0909.2348}}].

\bibitem{Boliev:1995xz}
M.~M. Boliev {\em et.~al.}, {\it {Search for supersymmetric dark matter with
  Baksan underground telescope}},  {\em Nucl. Phys. Proc. Suppl.} {\bf 48}
  (1996) 83--86.

\bibitem{Barger:2001ur}
V.~D. Barger, F.~Halzen, D.~Hooper, and C.~Kao, {\it {Indirect search for
  neutralino dark matter with high energy neutrinos}},  {\em Phys. Rev.} {\bf
  D65} (2002) 075022, [\href{http://xxx.lanl.gov/abs/hep-ph/0105182}{{\tt
  hep-ph/0105182}}].

\bibitem{Ackermann:2005fr}
{\bf AMANDA} Collaboration, M.~Ackermann {\em et.~al.}, {\it {Limits to the
  muon flux from neutralino annihilations in the Sun with the AMANDA
  detector}},  {\em Astropart. Phys.} {\bf 24} (2006) 459--466,
  [\href{http://xxx.lanl.gov/abs/astro-ph/0508518}{{\tt astro-ph/0508518}}].

\bibitem{Halzen:2005ar}
F.~Halzen and D.~Hooper, {\it {Prospects for detecting dark matter with
  neutrino telescopes in light of recent results from direct detection
  experiments}},  {\em Phys. Rev.} {\bf D73} (2006) 123507,
  [\href{http://xxx.lanl.gov/abs/hep-ph/0510048}{{\tt hep-ph/0510048}}].

\bibitem{Abbasi:2009uz}
{\bf ICECUBE} Collaboration, R.~Abbasi {\em et.~al.}, {\it {Limits on a muon
  flux from neutralino annihilations in the Sun with the IceCube 22-string
  detector}},  {\em Phys. Rev. Lett.} {\bf 102} (2009) 201302,
  [\href{http://xxx.lanl.gov/abs/0902.2460}{{\tt arXiv:0902.2460}}].

\bibitem{Loucatos:2010zz}
{\bf ANTARES} Collaboration, S.~Loucatos, {\it {Indirect search for dark matter
  with the ANTARES neutrino telescope}},  {\em AIP Conf. Proc.} {\bf 1200}
  (2010) 989--992.

\bibitem{Mena:2007ty}
O.~Mena, S.~Palomares-Ruiz, and S.~Pascoli, {\it {Reconstructing WIMP
  properties with neutrino detectors}},  {\em Phys. Lett.} {\bf B664} (2008)
  92--96, [\href{http://xxx.lanl.gov/abs/0706.3909}{{\tt arXiv:0706.3909}}].

\bibitem{Bueno:2004dv}
A.~Bueno, R.~Cid, S.~Navas, D.~Hooper, and T.~Weiler, {\it {Indirect detection
  of dark matter WIMPs in a liquid argon TPC}},  {\em JCAP} {\bf 0501} (2005)
  001, [\href{http://xxx.lanl.gov/abs/hep-ph/0410206}{{\tt hep-ph/0410206}}].

\bibitem{Kumar:2011hi}
J.~Kumar, J.~G. Learned, M.~Sakai, and S.~Smith, {\it {Dark Matter Detection
  With Electron Neutrinos in Liquid Scintillation Detectors}},
  \href{http://xxx.lanl.gov/abs/1103.3270}{{\tt arXiv:1103.3270}}.

\bibitem{Hirata:1987hu}
{\bf KAMIOKANDE-II} Collaboration, K.~Hirata {\em et.~al.}, {\it {Observation
  of a Neutrino Burst from the Supernova SN 1987a}},  {\em Phys. Rev. Lett.}
  {\bf 58} (1987) 1490--1493.

\bibitem{Fukuda:1998mi}
{\bf Super-Kamiokande} Collaboration, Y.~Fukuda {\em et.~al.}, {\it {Evidence
  for oscillation of atmospheric neutrinos}},  {\em Phys. Rev. Lett.} {\bf 81}
  (1998) 1562--1567, [\href{http://xxx.lanl.gov/abs/hep-ex/9807003}{{\tt
  hep-ex/9807003}}].

\bibitem{Ahmad:2002jz}
{\bf SNO} Collaboration, Q.~R. Ahmad {\em et.~al.}, {\it {Direct evidence for
  neutrino flavor transformation from neutral-current interactions in the
  Sudbury Neutrino Observatory}},  {\em Phys. Rev. Lett.} {\bf 89} (2002)
  011301, [\href{http://xxx.lanl.gov/abs/nucl-ex/0204008}{{\tt
  nucl-ex/0204008}}].

\bibitem{:2008ee}
{\bf KamLAND} Collaboration, S.~Abe {\em et.~al.}, {\it {Precision Measurement
  of Neutrino Oscillation Parameters with KamLAND}},  {\em Phys. Rev. Lett.}
  {\bf 100} (2008) 221803, [\href{http://xxx.lanl.gov/abs/0801.4589}{{\tt
  arXiv:0801.4589}}].

\bibitem{Arpesella:2008mt}
{\bf The Borexino} Collaboration, C.~Arpesella {\em et.~al.}, {\it {Direct
  Measurement of the Be-7 Solar Neutrino Flux with 192 Days of Borexino Data}},
   {\em Phys. Rev. Lett.} {\bf 101} (2008) 091302,
  [\href{http://xxx.lanl.gov/abs/0805.3843}{{\tt arXiv:0805.3843}}].

\bibitem{Ahn:2006zza}
{\bf K2K Collaboration} Collaboration, M.~Ahn {\em et.~al.}, {\it {Measurement
  of Neutrino Oscillation by the K2K Experiment}},  {\em Phys.Rev.} {\bf D74}
  (2006) 072003, [\href{http://xxx.lanl.gov/abs/hep-ex/0606032}{{\tt
  hep-ex/0606032}}].

\bibitem{Adamson:2008zt}
{\bf MINOS} Collaboration, P.~Adamson {\em et.~al.}, {\it {Measurement of
  Neutrino Oscillations with the MINOS Detectors in the NuMI Beam}},  {\em
  Phys. Rev. Lett.} {\bf 101} (2008) 131802,
  [\href{http://xxx.lanl.gov/abs/0806.2237}{{\tt arXiv:0806.2237}}].

\bibitem{GonzalezGarcia:2007ib}
M.~C. Gonzalez-Garcia and M.~Maltoni, {\it {Phenomenology with Massive
  Neutrinos}},  {\em Phys. Rept.} {\bf 460} (2008) 1--129,
  [\href{http://xxx.lanl.gov/abs/0704.1800}{{\tt arXiv:0704.1800}}].

\bibitem{Strumia:2006db}
A.~Strumia and F.~Vissani, {\it {Neutrino masses and mixings and \ldots}},
  \href{http://xxx.lanl.gov/abs/hep-ph/0606054}{{\tt hep-ph/0606054}}.

\bibitem{Itow:2001ee}
{\bf The T2K} Collaboration, Y.~Itow {\em et.~al.}, {\it {The JHF-Kamioka
  neutrino project}},  \href{http://xxx.lanl.gov/abs/hep-ex/0106019}{{\tt
  hep-ex/0106019}}.

\bibitem{Autiero:2007zj}
D.~Autiero, J.~Aysto, A.~Badertscher, L.~B. Bezrukov, J.~Bouchez, {\em
  et.~al.}, {\it {Large underground, liquid based detectors for astro-particle
  physics in Europe: Scientific case and prospects}},  {\em JCAP} {\bf 0711}
  (2007) 011, [\href{http://xxx.lanl.gov/abs/0705.0116}{{\tt
  arXiv:0705.0116}}].

\bibitem{Raby:2008pd}
S.~Raby, T.~Walker, K.~Babu, H.~Baer, A.~Balantekin, {\em et.~al.}, {\it {DUSEL
  Theory White Paper}},  \href{http://xxx.lanl.gov/abs/0810.4551}{{\tt
  arXiv:0810.4551}}.

\bibitem{Lesko:2008zza}
K.~T. Lesko, {\it {The US National Science Foundation's deep underground
  laboratory at Homestake - DUSEL}},  {\em J.Phys.Conf.Ser.} {\bf 120} (2008)
  052011.

\bibitem{Raychaudhuri:2008zz}
{\bf INO} Collaboration, A.~Raychaudhuri, {\it {Status of INO}},  {\em PoS}
  {\bf NUFACT08} (2008) 045.

\bibitem{Datar:2009zz}
V.~Datar, S.~Jena, S.~Kalmani, N.~Mondal, P.~Nagaraj, {\em et.~al.}, {\it
  {Development of glass resistive plate chambers for INO experiment}},  {\em
  Nucl.Instrum.Meth.} {\bf A602} (2009) 744--748.

\bibitem{Rubbia:2010zz}
{\bf LAGUNA Collaboration} Collaboration, A.~Rubbia, {\it {The LAGUNA design
  study: Towards giant liquid based underground detectors for neutrino physics
  and astrophysics and proton decay searches}},  {\em Acta Phys.Polon.} {\bf
  B41} (2010) 1727--1732.

\bibitem{Pati:1973rp}
J.~C. Pati and A.~Salam, {\it {Is Baryon Number Conserved?}},  {\em Phys. Rev.
  Lett.} {\bf 31} (1973) 661--664.

\bibitem{Rubbia:2010fm}
A.~Rubbia, {\it {A CERN-based high-intensity high-energy proton source for long
  baseline neutrino oscillation experiments with next- generation large
  underground detectors for proton decay searches and neutrino physics and
  astrophysics}},  \href{http://xxx.lanl.gov/abs/1003.1921}{{\tt
  arXiv:1003.1921}}.

\bibitem{Barger:2007yw}
V.~Barger, M.~Bishai, D.~Bogert, C.~Bromberg, A.~Curioni, {\em et.~al.}, {\it
  {Report of the US long baseline neutrino experiment study}},
  \href{http://xxx.lanl.gov/abs/0705.4396}{{\tt arXiv:0705.4396}}.

\bibitem{Bandyopadhyay:2007kx}
{\bf ISS Physics Working Group} Collaboration, A.~Bandyopadhyay {\em et.~al.},
  {\it {Physics at a future Neutrino Factory and super-beam facility}},
  \href{http://xxx.lanl.gov/abs/0710.4947}{{\tt 0710.4947}}.

\bibitem{Wurm:2011zn}
{\bf LENA} Collaboration, M.~Wurm {\em et.~al.}, {\it {The next-generation
  liquid-scintillator neutrino observatory LENA}},
  \href{http://xxx.lanl.gov/abs/1104.5620}{{\tt arXiv:1104.5620}}.

\bibitem{ThesisLaing}
A.~Laing, {\em Optimization of Detectors for the Golden Channel at a Neutrino
  Factory}.
\newblock PhD thesis, Glasgow university, 2010.

\bibitem{Bari:2003bt}
G.~Bari {\em et.~al.}, {\it {Analysis of the performance of the MONOLITH
  prototype}},  {\em Nucl. Instrum. Meth.} {\bf A508} (2003) 170--174.

\bibitem{crubbia}
C.~Rubbia, {\it The liquid-argon time projection chamber: a new concept for
  neutrino detector},  Tech. Rep. CERN-EP/77-08, 1977.

\bibitem{Amerio:2004ze}
{\bf ICARUS} Collaboration, S.~Amerio {\em et.~al.}, {\it {Design, construction
  and tests of the ICARUS T600 detector}},  {\em Nucl. Instrum. Meth.} {\bf
  A527} (2004) 329--410.

\bibitem{Varanini:2009zz}
F.~Varanini, {\it {Preliminary experimental results from the ICARUS test
  facility at INFN-LNL}},  {\em Nucl. Phys. Proc. Suppl.} {\bf 197} (2009)
  313--316.

\bibitem{Menegolli:2010zz}
{\bf ICARUS} Collaboration, A.~Menegolli, {\it {Status of the ICARUS T600
  detector at the LNGS}},  {\em J. Phys. Conf. Ser.} {\bf 203} (2010) 012107.

\bibitem{Rubbia:2004tz}
A.~Rubbia, {\it {Experiments for CP-violation: A giant liquid argon
  scintillation, Cerenkov and charge imaging experiment?}},
  \href{http://xxx.lanl.gov/abs/hep-ph/0402110}{{\tt hep-ph/0402110}}.

\bibitem{laguna}
``Laguna - large apparatus studying grand unification and neutrino
  astrophysics.''
\newblock {\tt http://www.laguna-science.eu/}.

\bibitem{lbne}
``Lbne - long-baseline neutrino experiment.''
\newblock {\tt http://lbne.fnal.gov/}.

\bibitem{Barger:2006kp}
V.~Barger, P.~Huber, D.~Marfatia, and W.~Winter, {\it {Upgraded experiments
  with super neutrino beams: Reach versus Exposure}},  {\em Phys. Rev.} {\bf
  D76} (2007) 031301, [\href{http://xxx.lanl.gov/abs/hep-ph/0610301}{{\tt
  hep-ph/0610301}}].

\bibitem{Cervera:2000vy}
A.~Cervera, F.~Dydak, and J.~Gomez~Cadenas, {\it {A large magnetic detector for
  the neutrino factory}},  {\em Nucl. Instrum. Meth.} {\bf A451} (2000)
  123--130.

\bibitem{Cervera:2010rz}
A.~Cervera, A.~Laing, J.~Martin-Albo, and F.~Soler, {\it {Performance of the
  MIND detector at a Neutrino Factory using realistic muon reconstruction}},
  {\em Nucl.Instrum.Meth.} {\bf A624} (2010) 601--614,
  [\href{http://xxx.lanl.gov/abs/1004.0358}{{\tt arXiv:1004.0358}}].

\bibitem{Geer:1997iz}
S.~Geer, {\it {Neutrino beams from muon storage rings: Characteristics and
  physics potential}},  {\em Phys. Rev.} {\bf D57} (1998) 6989--6997,
  [\href{http://xxx.lanl.gov/abs/hep-ph/9712290}{{\tt hep-ph/9712290}}].

\bibitem{DeRujula:1998hd}
A.~De~Rujula, M.~B. Gavela, and P.~Hernandez, {\it {Neutrino oscillation
  physics with a neutrino factory}},  {\em Nucl. Phys.} {\bf B547} (1999)
  21--38, [\href{http://xxx.lanl.gov/abs/hep-ph/9811390}{{\tt
  hep-ph/9811390}}].

\bibitem{ids}
``International design study of the neutrino factory.''
\newblock {\tt http://www.ids-nf.org}.

\bibitem{Abe:2007bi}
{\bf ISS Detector Working Group} Collaboration, T.~Abe {\em et.~al.}, {\it
  {Detectors and flux instrumentation for future neutrino facilities}},  {\em
  JINST} {\bf 4} (2009) T05001, [\href{http://xxx.lanl.gov/abs/0712.4129}{{\tt
  arXiv:0712.4129}}].

\bibitem{tifr-ino}
``India-based neutrino observatory.''
\newblock {\tt http://www.ino.tifr.res.in/ino/}.

\bibitem{Agarwalla:2009xc}
S.~K. Agarwalla, {\em {Some aspects of neutrino mixing and oscillations}}.
\newblock PhD thesis, 2009.
\newblock \href{http://xxx.lanl.gov/abs/0908.4267}{{\tt arXiv:0908.4267}}.

\bibitem{Agarwalla:2010hk}
S.~K. Agarwalla, P.~Huber, J.~Tang, and W.~Winter, {\it {Optimization of the
  Neutrino Factory, revisited}},  {\em JHEP} {\bf 1101} (2011) 120,
  [\href{http://xxx.lanl.gov/abs/1012.1872}{{\tt arXiv:1012.1872}}].

\bibitem{Michael:2008bc}
{\bf MINOS} Collaboration, D.~G. Michael {\em et.~al.}, {\it {The Magnetized
  steel and scintillator calorimeters of the MINOS experiment}},  {\em Nucl.
  Instrum. Meth.} {\bf A596} (2008) 190--228,
  [\href{http://xxx.lanl.gov/abs/0805.3170}{{\tt arXiv:0805.3170}}].

\bibitem{Adamson:2006xv}
P.~Adamson {\em et.~al.}, {\it {The MINOS calibration detector}},  {\em Nucl.
  Instrum. Meth.} {\bf A556} (2006) 119--133.

\bibitem{Cirelli:2005gh}
M.~Cirelli, N.~Fornengo, T.~Montaruli, I.~A. Sokalski, A.~Strumia, {\em
  et.~al.}, {\it {Spectra of neutrinos from dark matter annihilations}},  {\em
  Nucl.Phys.} {\bf B727} (2005) 99--138,
  [\href{http://xxx.lanl.gov/abs/hep-ph/0506298}{{\tt hep-ph/0506298}}].

\bibitem{Blennow:2007tw}
M.~Blennow, J.~Edsj{\"o}, and T.~Ohlsson, {\it {Neutrinos from WIMP
  annihilations using a full three-flavor Monte Carlo}},  {\em JCAP} {\bf 0801}
  (2008) 021, [\href{http://xxx.lanl.gov/abs/0709.3898}{{\tt
  arXiv:0709.3898}}].

\bibitem{Bahcall:2000nu}
J.~N. Bahcall, M.~H. Pinsonneault, and S.~Basu, {\it {Solar models: Current
  epoch and time dependences, neutrinos, and helioseismological properties}},
  {\em Astrophys. J.} {\bf 555} (2001) 990--1012,
  [\href{http://xxx.lanl.gov/abs/astro-ph/0010346}{{\tt astro-ph/0010346}}].

\bibitem{Gould}
A.~Gould, {\it {Cosmological density of WIMPs from solar and terrestrial
  annihilations}},  {\em Astrophys. J.} {\bf 388} (1992) 338--344.

\bibitem{Honda:2011nf}
M.~Honda, T.~Kajita, K.~Kasahara, and S.~Midorikawa, {\it {Improvement of low
  energy atmospheric neutrino flux calculation using the JAM nuclear
  interaction model}},  \href{http://xxx.lanl.gov/abs/1102.2688}{{\tt
  arXiv:1102.2688}}.

\bibitem{Blennow:2009pk}
M.~Blennow and E.~Fernandez-Martinez, {\it {Neutrino oscillation parameter
  sampling with MonteCUBES}},  {\em Comput. Phys. Commun.} {\bf 181} (2010)
  227--231, [\href{http://xxx.lanl.gov/abs/0903.3985}{{\tt arXiv:0903.3985}}].

\bibitem{Gelman:1992zz}
A.~Gelman and D.~B. Rubin, {\it {Inference from Iterative Simulation Using
  Multiple Sequences}},  {\em Statist. Sci.} {\bf 7} (1992) 457--472.

\end{thebibliography}

\providecommand{\href}[2]{#2}\begingroup\raggedright\endgroup

%
%
\end{document}